\documentclass[acmlarge]{acmart}

\usepackage{tabularx}
\usepackage{ragged2e}
\usepackage{colortbl}
\usepackage{xcolor}

\definecolor{sectionblue}{RGB}{0,90,160}

\usepackage{booktabs}
\usepackage{multirow}
\usepackage{array}
\usepackage{makecell}
\usepackage{graphicx}
\usepackage{rotating}

\newcolumntype{C}[1]{>{\centering\arraybackslash}p{#1}}
\newcolumntype{Y}{>{\centering\arraybackslash}X}

\AtBeginDocument{%
  }

\setcopyright{acmlicensed}
\copyrightyear{2026}
\acmYear{2026}
\acmDOI{}

\usepackage{xcolor}
\usepackage{listings}
\usepackage[most]{tcolorbox}

\usetikzlibrary{arrows.meta}

\definecolor{codepurple}{RGB}{154,43,157}
\definecolor{boxbackground}{RGB}{248,248,248}
\definecolor{boxborder}{RGB}{135,135,135}
\definecolor{traceheader}{RGB}{112,112,112}
\definecolor{linenumber}{RGB}{110,110,110}
\definecolor{treecolor}{RGB}{95,115,120}
\definecolor{summarybg}{RGB}{245,245,245}

\lstdefinestyle{agentpython}{
  language=Python,
  basicstyle=\ttfamily\fontsize{7.4}{9.2}\selectfont,
  numbers=left,
  numberstyle=
    \ttfamily\fontsize{6.8}{8}\selectfont\color{linenumber},
  numbersep=10pt,
  keywordstyle=\color{black},
  stringstyle=\color{codepurple},
  commentstyle=\color{gray},
  identifierstyle=\color{black},
  showstringspaces=false,
  keepspaces=true,
  columns=fullflexible,
  breaklines=true,
  breakatwhitespace=false,
  frame=none,
  xleftmargin=1.8em,
  aboveskip=0pt,
  belowskip=0pt
}

\AtBeginDocument{%
    \setlength{\textfloatsep}{10pt plus 0pt minus 0pt}%

    \setlength{\intextsep}{10pt plus 0pt minus 0pt}%

    \setlength{\dbltextfloatsep}{10pt plus 0pt minus 0pt}%
}

\begin{document}
\raggedbottom

\title{Log Analysis for LLM-Based Agents: A Lifecycle-Oriented Survey}

\author{Cong Wang}
\email{wangcongjcdd@tust.edu.cn}
\affiliation{%
  \institution{Tianjin University of Science and Technology}
  \state{Tianjin}
  \country{China}
}

\author{Yaowei Sun}
\email{sywei@tust.edu.cn}
\orcid{0009-0007-7613-7037}
\affiliation{%
  \institution{Tianjin University of Science and Technology}
  \state{Tianjin}
  \country{China}
}

\author{Guohua Xin}

\ifdefined\correspondingauthor
  \correspondingauthor
\else
  \authornote{Corresponding author.}
\fi

\affiliation{%
  \institution{Tianjin University}
  \city{Tianjin}
  \country{China}
}
\email{xinguohua@tju.edu.cn}

\author{Jiongchi Yu}
\affiliation{
  \institution{Singapore Management University}
 \state{Victoria Street}
 \country{Singapore}}
\email{jcyu.2022@smu.edu.sg}

\author{Shuangcheng Sang}
\email{2518223090@mail.tust.edu.cn}
\affiliation{%
  \institution{Tianjin University of Science and Technology}
  \state{Tianjin}
  \country{China}
}

\author{Guangquan Xu}
\email{losin@tju.edu.cn}
\affiliation{%
  \institution{Tianjin University}
  \state{Tianjin}
  \country{China}
}

\author{Maode Ma}
\email{mamaode@suat-sz.edu.cn}
\affiliation{%
  \institution{Shenzhen University of Advanced Technology}
  \state{Shenzhen}
  \country{China}
}

\renewcommand{\shortauthors}{Sun et al.}

\begin{abstract}
Logs generated by LLM-based agents support analysis of execution behavior beyond final outputs. Existing surveys give limited attention to dependencies across the lifecycle of these records. A lifecycle-oriented review of 90 core studies published between 2024 and August 2026 covers Agent Logging, Agent Log Structuring, and Agent Log Mining. Stage-specific comparisons reveal cross-stage dependencies and limitations. Detailed logs do not necessarily provide sufficient evidence. Structural uniformity does not guarantee semantic equivalence. Analytical claims require alignment with supporting evidence, reference criteria, and validation procedures. Representative open-source tools, public datasets, and future research directions are also summarized.
\end{abstract}


\maketitle

\section{Introduction}
\label{sec:introduction}

In recent years, LLM-based agents, including coding agents, research agents, and enterprise assistants, have become increasingly integrated into real-world workflows.
These systems often perform long-horizon tasks involving repeated model inference, tool invocation, and external interaction, which can incur substantial token consumption, latency, and operational cost~\cite{anjum2026promcp,kaplunovich2025optimizing,bai2026agentspendmoney}.
According to Deloitte's 2026 \textit{State of AI in the Enterprise} report,\footnote{\url{https://www.deloitte.com/content/dam/assets-zone3/us/en/docs/services/consulting/2026/state-of-ai-2026.pdf}}
74\% of surveyed companies expected to use agentic AI at least moderately within two years, whereas only 21\% reported having a mature governance model for autonomous agents.
The report also identifies real-time behavioral monitoring and audit trails covering complete sequences of agent actions as essential governance capabilities.
Together, these operational and governance needs highlight the importance of systematically capturing and analyzing runtime execution records to understand agent behavior and support reliable deployment.

\begin{figure*}[t]
    \centering
    \includegraphics[width=0.96\textwidth]{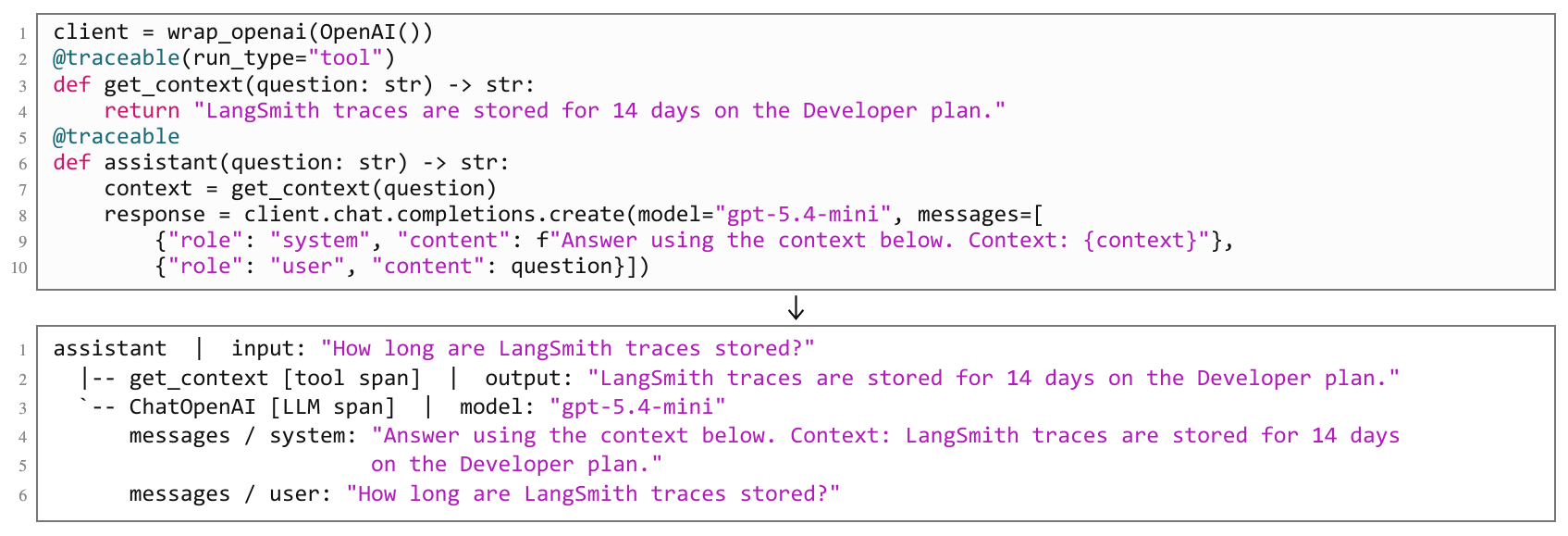}
    \caption{An example of agent instrumentation with LangSmith and the resulting execution trace.}
    \label{fig:langsmith-agent-logging}
\end{figure*}

Agent logs have been widely employed in agent monitoring and analysis because they record runtime information about intermediate decisions, tool interactions, and execution states that is often unavailable from final outputs alone \cite{yagoubi2026agentleak}.
They also provide an important data source for performance evaluation and operational monitoring of LLM-based agent systems \cite{kaplunovich2025optimizing}.
In general, agent logs are semi-structured execution records generated by instrumentation mechanisms, such as wrappers, decorators, and tracing hooks, during agent runtime.
For example, in Fig.~\ref{fig:langsmith-agent-logging}, execution spans are generated by the wrapper and decorators in the example code \cite{langsmithObservabilityDocs}.
The trace hierarchy and metadata, such as span types, model identifiers, and parent--child relations, are determined by the corresponding instrumentation framework and are therefore structured.
In contrast, runtime contents, such as user questions, retrieved context, and model messages, are largely unstructured because they contain task-specific information generated during agent execution.

In this context, automated agent log analysis has emerged to capture, organize, and analyze runtime execution records for understanding agent behavior.
In recent years, a growing body of work has investigated agent observability, execution assessment, safety analysis, replay, debugging, provenance, and failure diagnosis
\cite{dong2024agentops,ma2024agentboard,ruan2024toolemu,zhu2024redel,epperson2025interactive,souza2025provagent,zhang2025agentcauses}.
Recent studies have further explored LLM-based analysis and system-level tracing mechanisms for agent execution
\cite{rombaut2025watson,zhu2026raffles,zheng2025agentsight,seyedghorban2026observability}.
These growing research efforts show that agent-generated execution records have become an important basis for analyzing LLM-based agent systems.

\begin{figure*}[t]
    \centering
    \includegraphics[width=0.8\textwidth]{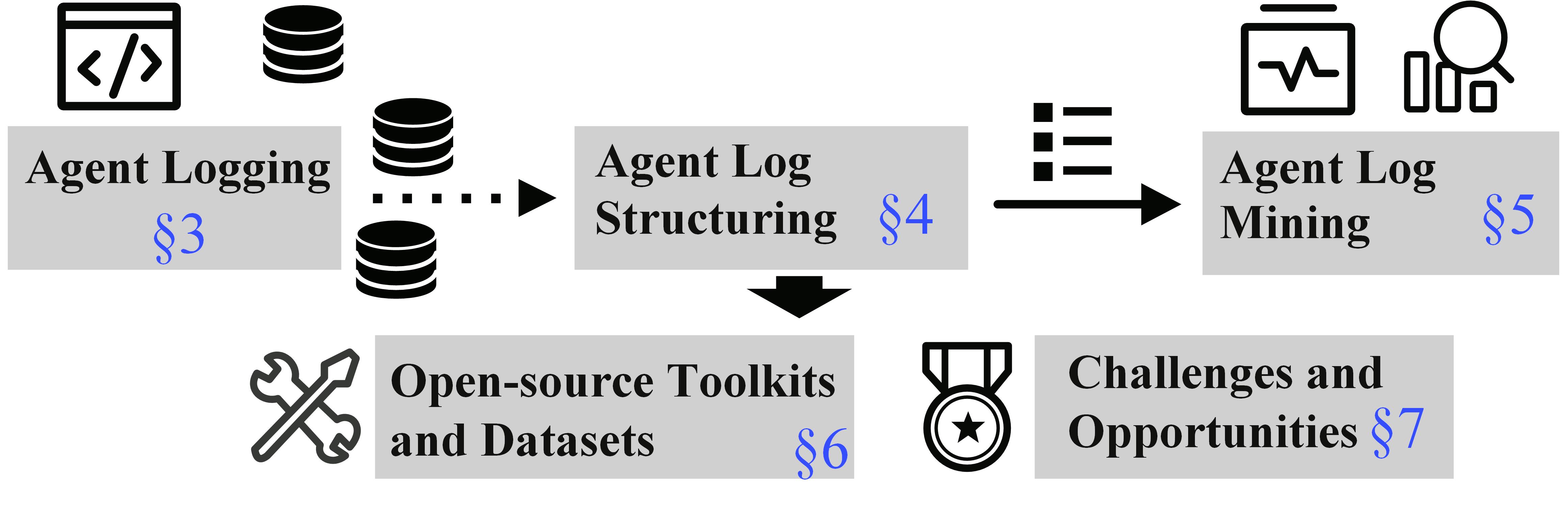}
    \caption{An overall framework of agent log analysis.}
    \label{fig:agent-log-analysis-taxonomy}
\end{figure*}

Several surveys and reviews have examined topics closely related to agent log analysis.
He et al.~\cite{he2021survey} systematically reviewed conventional software-log analysis, while recent studies have surveyed agent operations and observability \cite{dong2024agentops,wang2025agentopssurvey}, LLM-agent evaluation \cite{yehudai2026evaluationsurvey}, evidence tracing and execution provenance \cite{wang2026tracestotrust}, and trajectory analysis for failure attribution and system enhancement \cite{wang2026surveytrajectory}.
These studies provide important perspectives on software logs, agent operation, evaluation, provenance, and trajectory analysis.
However, they primarily organize the literature around conventional software logs or particular operational and analytical objectives rather than around the lifecycle of agent-generated execution records.
Consequently, research on how runtime records are generated, transformed into analyzable representations, and subsequently used by heterogeneous analysis tasks remains fragmented across these areas.
Therefore, our work provides a lifecycle-oriented survey of agent log analysis by treating persistent agent-generated execution records as the primary research object and organizing existing studies into three stages: Agent Logging, Agent Log Structuring, and Agent Log Mining. Therefore, our work provides a lifecycle-oriented survey of agent log analysis by treating persistent agent-generated execution records as the primary research object and organizing existing studies into three stages: Agent Logging, Agent Log Structuring, and Agent Log Mining.
The overall framework of agent log analysis is illustrated in Fig.~\ref{fig:agent-log-analysis-taxonomy}.
Furthermore, we systematically examine how existing approaches address these three stages and how execution records evolve across them.
Overall, unlike previous surveys, we focus on the lifecycle of agent-generated execution records from their generation to their transformation and downstream analysis.
Additionally, we establish stage-specific comparison criteria to characterize and compare existing approaches.
More importantly, we synthesize cross-study findings to reveal common limitations and identify future research directions.
Next, the main work and contributions of this article are summarized below.

\begin{itemize}
    \item \textbf{Systematic lifecycle taxonomy and review.}
    We organize automated agent log analysis into three stages: Agent Logging, Agent Log Structuring, and Agent Log Mining.
    Importantly, we establish stage-specific comparison criteria to characterize existing approaches at a fine-grained level.
    Then, we systematically review 90 core studies published from 2024 to 2026 and examine how agent-generated execution records are captured, structured, and analyzed across these stages.

    \item \textbf{Cross-study findings and discussion.}
    We summarize the methodological characteristics of existing approaches across the three stages.
    More importantly, we synthesize cross-study findings on their capabilities and limitations and reveal dependencies and gaps across logging, structuring, and mining that are difficult to expose from individual task-oriented studies.

    \item \textbf{More forward-looking challenges and opportunities.}
    Based on the reviewed literature and cross-study findings, we summarize current pressing issues and discuss future research opportunities for agent log analysis.
    In particular, we identify directions toward more complete, portable, verifiable, scalable, and analytically reliable analysis of agent-generated execution records.
\end{itemize}

The rest of the article is structured as follows.
Section~\ref{sec:methodology} describes the process of selecting papers for this survey.
Section~\ref{Agent Logging} reviews agent logging approaches and the corresponding comparison criteria.
Section~\ref{sec:agent-log-structuring} reviews approaches for transforming agent logs into structured execution representations.
Section~\ref{sec:agent-log-mining} reviews major agent log mining tasks and summarizes the corresponding cross-study findings.
Section~\ref{sec:Open-source-Toolkits-and-Datasets} presents representative open-source tools and public datasets.
Section~\ref{sec:challenges-and-opportunities} discusses current challenges and future research opportunities.
Section~\ref{sec:conclusion} concludes the article.

\section{Paper Selection}
\label{sec:methodology}

This section describes the workflow for conducting paper selection (the paper collection date is August 2026).

\textbf{Paper Collection.}
We searched several popular online digital libraries, including IEEE Xplore, ACM Digital Library, SpringerLink, Elsevier/ScienceDirect, and Wiley Online Library, using the following keywords: ``agent log,'' ``agent logging,'' ``agent log structuring,'' ``agent trajectory,'' ``agent observability,'' ``agent provenance,'' ``agent log + anomaly detection,'' ``agent log + failure prediction,'' and ``agent log + failure diagnosis.''
Given that systematic research on execution records generated by LLM-based agent systems has emerged only recently, we limited the publication period to studies published from 2023 to August 2026.
However, these search terms are also widely used in adjacent agent-related research, such as context processing, memory management, trajectory-based evaluation, and general agent observability, where execution records may be involved without being the primary object of generation, structuring, or analysis.
To improve the precision of our collection, we therefore mainly focused on regular papers published in major conferences and journals across relevant domains, including ICSE, FSE, ASE, TSE, TOSEM, EMSE, SOSP, OSDI, ATC, NSDI, TDSC, DSN, ASPLOS, ICML, ICLR, USENIX Security, and CCS.
These venues cover major research communities related to software engineering, systems, artificial intelligence, and security, providing a broad basis for collecting recent studies on agent log analysis.
After removing duplicate records retrieved from different digital libraries and venues, our initial collection contained 117 unique candidate papers, forming the starting point for the subsequent filtering process.

\textbf{Paper Filtering.}
Due to the broad scope of the initially collected literature, it is necessary to refine our selection of papers for inclusion in the main body of this survey.
To achieve this, we established three specific criteria for the filtering process:
(1) The studied system must contain one or more LLM-based agents, and the study must directly concern execution records generated during agent runtime or structured representations derived from such records. Studies in which trajectories are merely retained as incidental experimental artifacts or working context, as well as studies in which LLM-based agents analyze logs generated by conventional software systems, are excluded;
(2) Papers must make a substantive contribution to agent logging, agent log structuring, or agent log mining. For mining studies, execution records need not be the sole source of evidence or the ultimate property of interest, but they must constitute a substantive basis of the principal analytical process rather than auxiliary experimental information;
(3) Papers must provide sufficient methodological and empirical details to support the comparative analysis in this survey, including the studied problem, input execution records, principal output, core mechanism, and evaluation setting. Publication type and length are used as indicators of methodological depth rather than standalone exclusion rules. Accordingly, short abstracts, surveys, industrial documentation, and papers that only introduce tools or datasets are excluded when they do not provide sufficient technical and empirical detail.

After applying these criteria, we further conducted complementary venue-level screening in major conferences and journals of relevant domains to reduce omissions caused by keyword-based retrieval.
In addition, considering that several newly available studies had not yet been covered by the preceding collection, we manually retrieved eligible recent papers as a supplement.
Finally, we obtained 90 core studies, which form the primary literature corpus reviewed in Sections~\ref{Agent Logging}--\ref{sec:agent-log-mining}.

\begin{figure*}[t]
\centering
\includegraphics[width=0.8\textwidth]{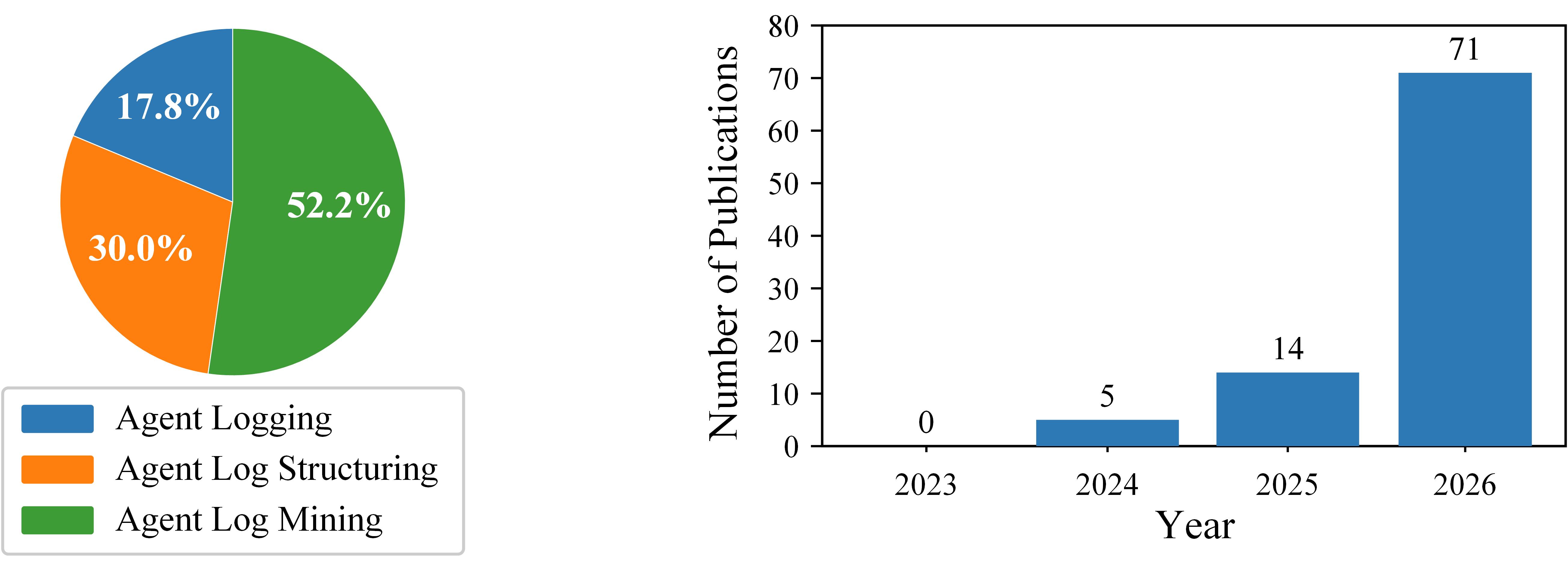}
\caption{Paper distribution on each research topic and the associated publication numbers from 2023 to 2026.}
\label{fig:methodology-distribution}
\end{figure*}

Figure~\ref{fig:methodology-distribution} summarizes the distribution of the reviewed studies by research stage and publication year.
Each study is assigned to a single stage according to its primary contribution, even when it spans multiple stages.

For further literature and resource discovery, readers may consult community-curated collections such as Awesome LLM Agent Papers\footnote{\url{https://github.com/js-lee-AI/awesome-llm-agent-papers}},
Awesome Agent Observability\footnote{\url{https://github.com/danielt69/awesome-agent-observability}},
Awesome Auditable AI\footnote{\url{https://github.com/yzhao062/awesome-auditable-ai}},
and Awesome-Agent-Trajectory\footnote{\url{https://github.com/YintongHuo/awesome-agent-trajectory}}.
These collections serve as supplementary resources; the comparative analysis remains based on the 90 core studies selected above.

\section{Agent Logging}
\label{Agent Logging}

Agent logging is the process of capturing runtime information generated during LLM agent execution and organizing it into agent logs. Unlike conventional software logs, agent logs need to record agent-specific runtime information (e.g., model interactions, tool use, and inter-agent communication) generated during agent execution. It has attracted growing attention from both academia \cite{dong2024agentops,alsayyad2026agenttrace,bindschaedler2026flightrecorder,aryan2026semanticprovenance,mishra2026observability,zheng2025agentsight,rorseth2025ladybug,anjum2026promcp,zhu2024redel,kaplunovich2025optimizing,hu2026oxygent} and industry \cite{openllmetryIntroduction,langsmithObservabilityDocs,langfuseObservabilityDocs,phoenixTracingDocs} because the generated agent logs serve as the input for subsequent agent log structuring and mining tasks.

\subsection{Comparison Criteria for Agent Logging Approaches}
\label{sec:logging-approaches}

In this section, we propose a uniform set of comparison criteria for reviewing and comparing different agent logging approaches (see Table~\ref{tab:logging-criteria}). This set of criteria characterizes agent logging approaches from three dimensions: \textbf{Where-to-log}, \textbf{What-to-log}, and \textbf{How-to-log}. These three dimensions contain 10 metrics, which reflect the similarities and differences among different agent logging approaches from different perspectives.

\begin{table*}[t]
\centering
\caption{A Uniform Set of Comparison Criteria for Agent Logging Approaches}
\label{tab:logging-criteria}
\begingroup
\renewcommand{\arraystretch}{1.12}
\setlength{\tabcolsep}{4pt}
\scriptsize
\begin{tabularx}{\textwidth}{
|>{\raggedright\arraybackslash}p{0.26\textwidth}|
>{\raggedright\arraybackslash}X|
}
\hline
\centering\textbf{Dimension} &
\centering\arraybackslash\textbf{Metrics} \\
\hline

\textbf{Where-to-log.}
This dimension describes where agent runtime information is recorded in agent logs. Agent execution can be logged at different runtime locations, and different approaches may cover one or multiple log sources. We characterize the log source locations through three metrics.
&
\textbf{(1) Execution (EX).}
This metric indicates whether a logging approach captures agent execution at the run or workflow level. Approaches that associate multiple events or steps with the same execution provide execution-level logging, whereas isolated model, tool, or function invocations alone are insufficient. Such logging preserves execution continuity across agent workflows.
\par\smallskip\hrule\smallskip
\textbf{(2) Invocation (IV).}
This metric indicates whether a logging approach captures individual calls to models or tools during execution. Approaches that record these call boundaries provide invocation-level logging, whereas run-level events or cross-agent messages alone are insufficient. Such logging provides fine-grained information about runtime operations.
\par\smallskip\hrule\smallskip
\textbf{(3) Communication (CM).}
This metric indicates whether a logging approach captures interactions across agent or system-component boundaries. Recording inter-agent messages, delegation handoffs, or protocol exchanges qualifies as communication-level logging, whereas local model, tool, or function calls alone do not. This metric reflects whether logging covers interactions beyond local execution boundaries.
\\
\hline

\textbf{What-to-log.}
This dimension describes the runtime information recorded in agent logs. Different approaches may record different types of information, including execution context and reasoning information. We characterize the recorded information through five metrics.
&
\textbf{(4) Execution Context (EC).}
This metric indicates whether a logging approach records execution context associated with participating agents or workflow entities. Approaches that retain persistent or evolving states together with entity identities or relations provide execution context, whereas static identifiers or isolated event attributes alone are insufficient. Such context links execution events with their corresponding entities and states.
\par\smallskip\hrule\smallskip
\textbf{(5) Input and Result (IOR).}
This metric indicates whether a logging approach records the inputs and results of agent operations. Such information may include prompts and responses, tool arguments and returned results, or messages and outputs. Operation identifiers, timestamps, or execution status alone are insufficient without the corresponding input and result content.
\par\smallskip\hrule\smallskip
\textbf{(6) Runtime Metadata (RM).}
This metric indicates whether a logging approach records quantitative runtime information, including timestamps, latency, token usage, resource consumption, or monetary cost. Such metadata characterizes the temporal and computational aspects of agent execution.
\par\smallskip\hrule\smallskip
\textbf{(7) Reasoning (RS).}
This metric indicates whether a logging approach records reasoning-related information explicitly exposed during agent execution. Such information may include plans or rationales that explain how an agent selects or revises its actions. Prompts, responses, or observable actions alone are insufficient without explicitly recorded reasoning information.
\par\smallskip\hrule\smallskip
\textbf{(8) Permission Information (PI).}
This metric indicates whether a logging approach records permission-related information associated with agent actions, including granted permissions, access scopes, authorization constraints, or execution privileges. Agent identities or interaction relations alone are insufficient unless the corresponding permission information that determines whether an action is allowed is explicitly recorded.
\\
\hline

\textbf{How-to-log.}
This dimension describes how runtime information is captured to generate agent logs. Existing approaches mainly differ in whether runtime information is captured through instrumentation integrated into the agent runtime or framework, or through interception at external execution boundaries. We characterize the logging mechanisms through two metrics.
&
\textbf{(9) Integrated Instrumentation (II).}
This metric indicates whether a logging approach captures runtime information through mechanisms integrated into the agent runtime, framework, or workflow. Approaches using event mechanisms, wrappers, callbacks, hooks, or tracing APIs within the execution environment are considered integrated instrumentation, whereas purely external collection is insufficient. Such instrumentation enables direct access to agent-specific runtime information.
\par\smallskip\hrule\smallskip
\textbf{(10) Boundary Interception (BI).}
This metric indicates whether a logging approach captures runtime information by intercepting interactions at tool, gateway, protocol, or system boundaries. Unlike integrated instrumentation, these approaches collect logs from observable interfaces without requiring internal runtime integration.
\\
\hline

\end{tabularx}
\endgroup
\end{table*}

\subsection{Taxonomy of Agent Logging Approaches}
\label{sec:logging-taxonomy}

We use these three logging dimensions as the basis for reviewing existing agent logging approaches. We review agent logging approaches along three dimensions: \textbf{Where-to-log}, \textbf{What-to-log}, and \textbf{How-to-log}. Here, we analyze existing agent logging studies based on the comparison criteria proposed in Section~\ref{sec:logging-approaches}, and Table~\ref{tab:logging_comparison} provides a uniform view of the differences among existing approaches across these three dimensions. Next, we review existing agent logging approaches from these three dimensions in detail.

\begin{table*}[t]
\centering
\caption{Comparison of Agent Logging Studies}
\label{tab:logging_comparison}

\begingroup
\renewcommand{\arraystretch}{1.15}
\setlength{\tabcolsep}{4pt}
\scriptsize

\resizebox{0.95\textwidth}{!}{
\begin{tabularx}{\textwidth}{
|>{\centering\arraybackslash}p{0.275\textwidth}|
>{\centering\arraybackslash}X|
>{\centering\arraybackslash}X|
>{\centering\arraybackslash}X|
>{\centering\arraybackslash}X|
>{\centering\arraybackslash}X|
>{\centering\arraybackslash}X|
>{\centering\arraybackslash}X|
>{\centering\arraybackslash}X|
>{\centering\arraybackslash}X|
>{\centering\arraybackslash}X|
>{\centering\arraybackslash}X|
>{\centering\arraybackslash}X|
}
\hline

\multirow{2}{*}{\textbf{Study}}
& \multicolumn{3}{c|}{\cellcolor{gray!25}\textbf{\textcolor{sectionblue}{\ref{sec:where-to-log}} Where-to-log}}
& \multicolumn{5}{c|}{\cellcolor{gray!25}\textbf{\textcolor{sectionblue}{\ref{sec:what-to-log}} What-to-log}}
& \multicolumn{2}{c|}{\cellcolor{gray!25}\textbf{\textcolor{sectionblue}{\ref{sec:how-to-log}} How-to-log}} \\

\cline{2-11}

& \textbf{EX}
& \textbf{IV}
& \textbf{CM}
& \textbf{EC}
& \textbf{IOR}
& \textbf{RM}
& \textbf{RS}
& \textbf{PI}
& \textbf{II}
& \textbf{BI} \\
\hline

Sequeira et al.~\cite{sequeira2026agentsentry}
& N & Y & N
& N & Y & N & N & N
& Y & Y \\
\hline

Dong et al.~\cite{dong2024agentops}
& Y & Y & N
& Y & Y & Y & Y & Y
& N & N \\
\hline

Zheng et al.~\cite{zheng2025agentsight}
& N & Y & N
& Y & Y & Y & N & N
& N & Y \\
\hline

AlSayyad et al.~\cite{alsayyad2026agenttrace}
& Y & Y & N
& N & Y & Y & Y & N
& Y & N \\
\hline

Pathak et al.~\cite{pathak2026governancetelemetry}
& N & N & Y
& Y & N & Y & N & Y
& Y & N \\
\hline

Sokolov et al.~\cite{sokolov2026ide}
& Y & Y & N
& N & Y & Y & N & N
& Y & N \\
\hline

Mishra et al.~\cite{mishra2026observability}
& N & Y & Y
& N & N & Y & N & Y
& N & Y \\
\hline

Seyedghorban et al.~\cite{seyedghorban2026observability}
& Y & Y & Y
& N & N & Y & N & N
& Y & N \\
\hline

Kaplunovich~\cite{kaplunovich2025optimizing}
& Y & Y & N
& Y & Y & Y & N & N
& Y & N \\
\hline

Hu et al.~\cite{hu2026oxygent}
& Y & Y & Y
& Y & Y & Y & Y & N
& Y & N \\
\hline

Anjum et al.~\cite{anjum2026promcp}
& N & Y & Y
& N & Y & Y & Y & N
& Y & N \\
\hline

Zhu et al.~\cite{zhu2024redel}
& Y & Y & Y
& Y & Y & Y & N & N
& Y & N \\
\hline

Vispute et al.~\cite{vispute2026reasoning}
& Y & Y & N
& Y & Y & Y & Y & Y
& Y & Y \\
\hline

Kaptein et al.~\cite{kaptein2026runtimegovernance}
& Y & Y & Y
& Y & Y & N & N & Y
& Y & N \\
\hline

Pasha et al.~\cite{pasha2026blockchain}
& Y & Y & Y
& Y & N & Y & Y & N
& Y & N \\
\hline

Zerhoudi et al.~\cite{zerhoudi2026agentsim}
& Y & Y & N
& N & Y & N & Y & N
& Y & N \\
\hline

\multicolumn{11}{|p{0.95\textwidth}|}{
Where-to-log: observation scope of the logging approach;
What-to-log: information retained in agent execution records;
How-to-log: mechanism used to capture execution evidence;
Y: Yes; N: No;
EX: Execution; IV: Invocation; CM: Communication;
EC: Execution Context; IOR: Input and Result; RM: Runtime Metadata;
RS: Reasoning; PI: Permission information;
II: Integrated Instrumentation; BI: Boundary Interception.
} \\
\hline

\end{tabularx}
}
\endgroup
\end{table*}

\subsubsection{Where-to-log}
\label{sec:where-to-log}
Agent execution can be observed from different runtime locations, and existing logging approaches differ in the scope of information they capture. Some approaches focus on individual runtime operations, while others extend observation to complete execution contexts or interactions across agent and system boundaries. These observation scopes reflect different logging capabilities and motivate the classification of \textbf{Where-to-log} into three locations: \textbf{Execution (EX)}, \textbf{Invocation (IV)}, and \textbf{Communication (CM)}.

\textbf{Execution-level logging (EX)} captures agent execution at the run or workflow level. ReDel organizes agent creation, state changes, messages, and model-related events within the execution of recursive multi-agent systems \cite{zhu2024redel}. AgentTrace associates method-, model-, and contextual events through shared trace contexts \cite{alsayyad2026agenttrace}, while OpenTelemetry-based approaches connect workflow, agent, model, and tool activities through trace propagation and parent--child span relations \cite{seyedghorban2026observability}. Other approaches explicitly represent execution sequences, such as Agent Execution Records that organize multiple execution steps and their associated evidence \cite{vispute2026reasoning}, or runtime contexts that preserve partial execution paths \cite{kaptein2026runtimegovernance}. These approaches extend logging from individual operations to correlated execution contexts, enabling multiple runtime activities to be associated with the same agent workflow.

\textbf{Invocation-level logging (IV)} captures individual execution invocations at explicit runtime boundaries (e.g., model calls, tool uses, function executions, or agent methods). ReDel exposes model and agent operations through its event-driven execution layer \cite{zhu2024redel}, while Kaplunovich integrates observability into model interaction points within a multi-agent code-generation workflow \cite{kaplunovich2025optimizing}. AgentTrace instruments public agent methods and related LLM interactions through runtime wrappers \cite{alsayyad2026agenttrace}, Sokolov et al. employ framework callbacks to capture node, model, and tool activities during IDE-triggered executions \cite{sokolov2026ide}, and Seyedghorban et al. represent model and tool operations using OpenTelemetry spans \cite{seyedghorban2026observability}. These approaches capture fine-grained runtime operations through explicit invocation boundaries.

\textbf{Communication-level logging (CM)} captures interactions across agent or system-component boundaries. ReDel records agent messages and dynamically formed agent relationships within recursive multi-agent execution \cite{zhu2024redel}. Mishra et al. preserve delegation identifiers and execution context across agents, tools, and external services \cite{mishra2026observability}, while Pathak et al. incorporate source and receiving agents together with lineage information into telemetry records \cite{pathak2026governancetelemetry}. Seyedghorban et al. represent agent-to-agent interactions through distributed tracing structures \cite{seyedghorban2026observability}, and ProMCP observes interactions across the Host--Client--Server path of MCP-based tool execution \cite{anjum2026promcp}. Such approaches require additional relational information to associate communication events with their participating entities across execution boundaries.

\begin{tcolorbox}[
    colback=summarybg,
    colframe=black,
    boxrule=0.5pt,
    arc=0pt,
    left=5pt,
    right=5pt,
    top=4pt,
    bottom=4pt
]
\textbf{Summary.}
\textbf{Where-to-log} defines the observation scope of agent logging by specifying the runtime locations at which execution information is captured. Existing approaches primarily focus on recording local execution activities and associating them with broader execution contexts through different logging scopes \cite{zhu2024redel,alsayyad2026agenttrace,seyedghorban2026observability,vispute2026reasoning}. Although some approaches extend logging across agent and system boundaries, such communication-level capture requires additional relational information to maintain cross-entity context \cite{mishra2026observability,pathak2026governancetelemetry,anjum2026promcp}. Therefore, the main distinction among existing approaches lies in how logging coverage expands from individual runtime operations to broader execution processes and interactions.
\end{tcolorbox}

\subsubsection{What-to-log}
\label{sec:what-to-log}

\textbf{What-to-log} characterizes the runtime information retained for interpreting agent execution, covering operational and semantic aspects through five metrics.

\textbf{Execution Context (EC)} records context associated with agents or workflow entities. ReDel retains agent identities, dynamic relationships, and event-specific information~\cite{zhu2024redel}, while AgentTrace links activities through identifiers and trace/span relationships within a common event structure~\cite{alsayyad2026agenttrace}. Sokolov et al. preserve context for node, model, and tool activities~\cite{sokolov2026ide}, whereas OpenTelemetry-based approaches associate operations through trace identities and parent--child span relations~\cite{seyedghorban2026observability}.

\textbf{Input and Result (IOR)} records operation inputs and corresponding outputs. Kaplunovich preserves prompts and model outputs throughout a multi-agent code-generation workflow~\cite{kaplunovich2025optimizing}, while AgentTrace records method arguments and return values~\cite{alsayyad2026agenttrace}.

\textbf{Runtime Metadata (RM)} records quantitative execution characteristics. ReDel retains token usage in its event stream~\cite{zhu2024redel}, while Kaplunovich records cost-related information across pipeline stages~\cite{kaplunovich2025optimizing}. ProMCP captures token usage and latency across MCP interaction stages~\cite{anjum2026promcp}, and Hu et al. derive runtime indicators from structured multi-agent execution events~\cite{hu2026oxygent}.

\textbf{Reasoning (RS)} records reasoning-related information explicitly exposed during execution. AgentTrace includes reasoning artifacts alongside operational and contextual events~\cite{alsayyad2026agenttrace}, while Vispute et al. organize intent, observation, inference, evidence, and plan revisions within Agent Execution Record steps~\cite{vispute2026reasoning}.

\textbf{Permission Information (PI)} records authorization-related information associated with agent actions. Mishra et al. link delegation context and authorization information to actions across agents and external tools~\cite{mishra2026observability}, while Pathak et al. retain source and receiving agents, lineage, and access-related attributes in telemetry~\cite{pathak2026governancetelemetry}. Kaptein et al. preserve execution paths and agent identities as context for evaluating runtime actions under predefined policies~\cite{kaptein2026runtimegovernance}.

\begin{tcolorbox}[
    colback=summarybg,
    colframe=black,
    boxrule=0.5pt,
    arc=0pt,
    left=5pt,
    right=5pt,
    top=4pt,
    bottom=4pt
]
\textbf{Summary.}
\textbf{What-to-log} characterizes the types of runtime information retained in agent logs. Existing approaches primarily focus on recording structural and operational information associated with agent activities, while some approaches extend recorded information to execution states, reasoning-related information, and permission-related context \cite{zhu2024redel,kaplunovich2025optimizing,alsayyad2026agenttrace,vispute2026reasoning,mishra2026observability}. These extensions enrich the information available for interpreting agent execution beyond individual runtime activities. Therefore, the main distinction among existing approaches lies in how recorded information expands from operational execution records to higher-level semantic and contextual information.
\end{tcolorbox}

\subsubsection{How-to-log}
\label{sec:how-to-log}
Agent logging approaches differ in how runtime information is captured and transformed into execution records. Existing approaches obtain execution information either through mechanisms integrated within agent execution environments or through observation at external execution boundaries. These capture mechanisms define \textbf{How-to-log} in agent logging, which can be characterized through two metrics: \textbf{Integrated Instrumentation (II)} and \textbf{Boundary Interception (BI)}.

\textbf{Integrated Instrumentation (II)} captures runtime information through mechanisms embedded in agent runtimes, workflows, or frameworks. ReDel emits and persists built-in and developer-defined events during recursive multi-agent execution~\cite{zhu2024redel}, while Kaplunovich embeds logging throughout a multi-agent code-generation pipeline~\cite{kaplunovich2025optimizing}. For reusable instrumentation, AgentTrace uses Python introspection, wrappers, and decorators to instrument agent methods and runtime interactions~\cite{alsayyad2026agenttrace}, whereas Sokolov et al. capture node, model, and tool activities through framework callbacks and wrappers during IDE-triggered execution~\cite{sokolov2026ide}. Seyedghorban et al. trace heterogeneous agent operations through a common OpenTelemetry abstraction~\cite{seyedghorban2026observability}; ProMCP instruments multiple MCP Host--Client--Server stages to collect protocol-level runtime information~\cite{anjum2026promcp}; and Pasha et al. capture agent-node outputs through workflow-level hooks~\cite{pasha2026blockchain}.

\textbf{Boundary Interception (BI)} captures runtime information at external tool, gateway, protocol, or system boundaries. AgentSight uses eBPF probes at stable SSL and kernel interfaces to monitor LLM interactions and system-level effects from outside the application~\cite{zheng2025agentsight}. Mishra et al. intercept delegated execution at a shared gateway between agents and external tools or services to associate actions across heterogeneous components~\cite{mishra2026observability}. Agent-Sentry intercepts execution before security-sensitive action--tool calls~\cite{sequeira2026agentsentry}, while Vispute et al. incorporate externally observable protocol and system operations into execution records~\cite{vispute2026reasoning}. These approaches extend observation beyond internally instrumented execution environments.

\begin{tcolorbox}[
    colback=summarybg,
    colframe=black,
    boxrule=0.5pt,
    arc=0pt,
    left=5pt,
    right=5pt,
    top=4pt,
    bottom=4pt
]
\textbf{Summary.}
\textbf{How-to-log} characterizes the mechanisms for transforming runtime information into agent logs. Existing approaches mainly rely on integrated instrumentation to capture runtime information within agent execution environments, while some approaches extend observation through boundary interception beyond internal runtimes \cite{zhu2024redel,alsayyad2026agenttrace,sokolov2026ide,seyedghorban2026observability,zheng2025agentsight}. Although boundary-based mechanisms broaden the observable execution scope, they require additional context attribution to associate external interactions with agent-level activities \cite{mishra2026observability,sequeira2026agentsentry,vispute2026reasoning}. Therefore, the main distinction among existing approaches lies in how logging mechanisms balance internal runtime integration with external observation coverage.
\end{tcolorbox}

\begin{figure*}[t]
\centering
\includegraphics[width=0.96\textwidth]{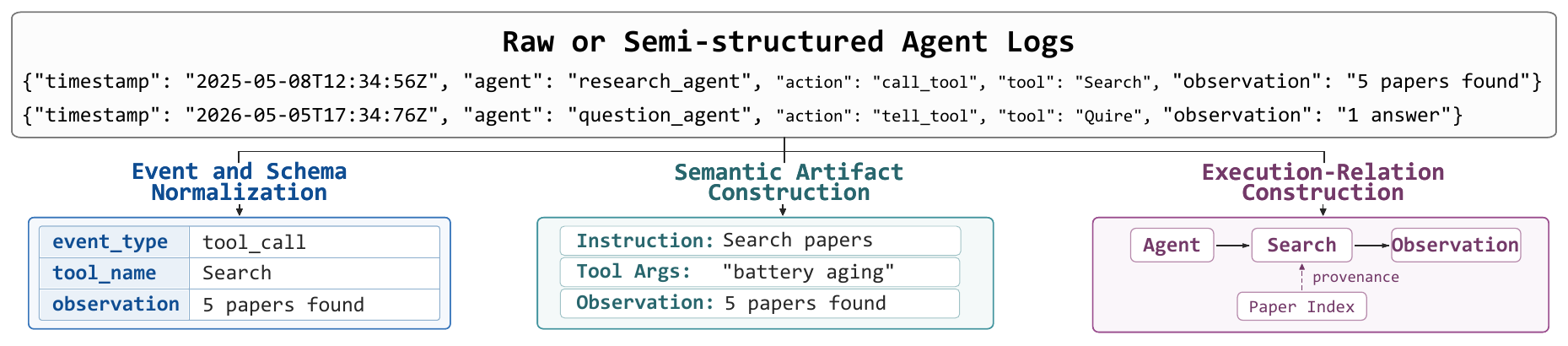}
\caption{Output-oriented categories of agent log structuring.}
\label{fig:agent-log-structuring}
\end{figure*}

\subsection{Discussion}
\label{sec:logging-discussion}

\textbf{Finding 1: Observation Scope.}
\textbf{\textit{Existing agent logging approaches provide extensive visibility into local execution activities, whereas interactions across agents and system components are less consistently captured.}}
Current approaches widely record individual model, tool, and method activities and associate them with the corresponding execution context (see Table~\ref{tab:logging_comparison}). However, observing an interaction across execution entities requires more than recording the activities at its two endpoints, since the participating entities and their relationships must also be preserved across execution boundaries. Some approaches have addressed this issue by recording agent messages, delegation contexts, lineage information, or distributed interaction traces~\cite{zhu2024redel,mishra2026observability,pathak2026governancetelemetry,seyedghorban2026observability}. Nevertheless, this information is not consistently retained across existing logging approaches. Consequently, an execution may provide detailed records of individual runtime activities while still lacking continuous visibility into how those activities are connected across interacting agents or components.

\textbf{Finding 2: Semantic Coverage.}
\textbf{\textit{Existing agent logs generally provide rich operational information, but reasoning and permission-related semantics remain less consistently recorded.}}
Inputs, outputs, runtime characteristics, and execution context are commonly retained in existing agent logs, whereas information explaining an agent's reasoning process or the permissions governing its actions appears in a smaller set of approaches (see Table~\ref{tab:logging_comparison}). This difference arises because operational information can often be obtained directly from observable runtime activities, while reasoning and permission semantics need to be explicitly exposed or associated with the corresponding action. Some approaches have begun to preserve reasoning artifacts or authorization-related context~\cite{alsayyad2026agenttrace,vispute2026reasoning,mishra2026observability,pathak2026governancetelemetry,kaptein2026runtimegovernance}. However, when such information is absent from the original logs, subsequent structuring can only construct or infer additional semantic artifacts from the recorded execution (see Section~\ref{sec:semantic-artifact-construction}). Therefore, detailed operational records do not necessarily imply equally complete semantic information about why an agent acted or under what permissions the action occurred.

\textbf{Finding 3: Logging Mechanism.}
\textbf{\textit{Agent logging is dominated by mechanisms integrated into execution environments, while external observation provides complementary coverage that is only rarely combined with rich internal execution context.}}
Most existing approaches embed logging mechanisms into agent runtimes, workflows, or frameworks, which allows runtime information to be captured close to the operations that generate agent-specific context (see Table~\ref{tab:logging_comparison}). In contrast, a smaller group observes interactions at tool, gateway, protocol, or system boundaries, allowing activities outside internally instrumented components to be captured~\cite{zheng2025agentsight,mishra2026observability,sequeira2026agentsentry,vispute2026reasoning}. However, such externally observed interactions usually contain less agent-level context and therefore require additional attribution or correlation, whereas internally integrated logging remains dependent on the execution environment that exposes this context. Only a few approaches combine information obtained from both sides of this boundary~\cite{sequeira2026agentsentry,vispute2026reasoning}. Overall, existing log generation strategies have not yet consistently combined rich agent-specific context with observation beyond the internally instrumented runtime.

\section{Agent Log Structuring}
\label{sec:agent-log-structuring}

Agent log structuring transforms previously emitted raw or semi-structured execution records into organized representations, including normalized events, semantic artifacts, and explicit execution relations. Because agent logs are heterogeneous across frameworks and often contain weakly structured content and implicit dependencies, existing approaches normalize records, extract semantic artifacts, or reconstruct execution relations \cite{song2026adp,barke2026agentrx,billiris2026foap,sriraman2026contextviewer,souza2025provagent,li2026delinearizing}. These representations provide analyzable execution evidence for downstream tasks such as failure diagnosis, agent execution assessment, and interactive inspection \cite{ou2025agentdiagnose,kim2025beyond,epperson2025interactive}.

\subsection{Comparison Criteria for Agent Log Structuring Approaches}
\label{sec:structuring_criteria}

In this section, we propose a uniform set of comparison criteria for reviewing and comparing different agent log structuring approaches (see Table~\ref{tab:structuring_criteria}). This set of criteria characterizes agent log structuring approaches in two dimensions: \textbf{Structuring Design} and \textbf{Structuring Ability}. These two dimensions contain six metrics, which reflect the similarities and differences among existing agent log structuring approaches from different perspectives.

\begin{table*}[t]
\centering
\caption{A Uniform Set of Comparison Criteria for Agent Log Structuring Approaches}
\label{tab:structuring_criteria}
\begingroup
\renewcommand{\arraystretch}{1.12}
\setlength{\tabcolsep}{4pt}
\scriptsize
\begin{tabularx}{\textwidth}{
|>{\raggedright\arraybackslash}p{0.275\textwidth}|
>{\raggedright\arraybackslash}X|
}
\hline
\multicolumn{1}{|c|}{\textbf{Dimensions}} &
\multicolumn{1}{c|}{\textbf{Metrics}} \\
\hline

\textbf{Structuring Design.} Agent log structuring approaches differ in how recorded agent logs are transformed, when the structured representation is constructed, and whether external knowledge is used during this process. We characterize these design differences through three metrics.
&
\textbf{(1) Structuring Mechanism (SM).} This metric indicates the principal mechanism used to construct the target structured representation. We categorize existing approaches as \textit{Rule-based}, \textit{LLM-based}, \textit{Human-involved}, or \textit{Learning-based}, according to whether the representation is primarily produced through predefined rules or mappings, pretrained LLMs, direct human intervention, or learned/data-driven models.
\par\smallskip\hrule\smallskip
\textbf{(2) Operation Mode (OM).} This metric indicates whether structuring occurs during or after agent execution. \textit{Online} approaches construct or incrementally update the representation while the agent execution is still ongoing, whereas \textit{Offline} approaches structure logs after the corresponding execution records have been generated. Interactive inspection of an already constructed representation is not considered online structuring.
\par\smallskip\hrule\smallskip
\textbf{(3) External Knowledge Utilization (EK).} This metric indicates whether information beyond the recorded agent logs is used to guide construction of the representation, such as domain knowledge, ontologies, external annotations, or reference structures. A predefined output schema alone is not considered external knowledge unless it introduces additional semantic or domain-specific information.
\\
\hline

\textbf{Structuring Ability.}
Structuring approaches differ in the types of agent executions and log records to which they have been applied, as well as in their ability to remain effective across different logs. We characterize these differences through three metrics.
&
\textbf{(4) Multi-Agent Structuring (MA).}
This metric indicates whether an approach has been applied to execution records generated by multi-agent systems. It characterizes the agent configuration of the evaluated logs rather than the potential extensibility of the approach to multi-agent settings.
\par\smallskip\hrule\smallskip
\textbf{(5) Cross-Log Transferability (CT).}
This metric indicates whether the core structuring approach remains effective across different agent logs, which may vary in systems, frameworks, tasks, workflows, environments, or logging formats. It therefore captures demonstrated applicability beyond the logs on which the approach was originally developed or evaluated.
\par\smallskip\hrule\smallskip
\textbf{(6) Heterogeneous Record Structuring (HR).}
This metric indicates whether an approach can jointly organize semantically distinct execution records, such as messages, model invocations, tool interactions, actions, observations, states, and workflow events. Support is identified when these record types are explicitly distinguished or associated in the resulting representation rather than treated as an undifferentiated sequence.
\\
\hline

\end{tabularx}
\endgroup
\end{table*}

\subsection{Taxonomy of Agent Log Structuring Approaches}
\label{sec:structuring-taxonomy}

We use the resulting structured representation as the basis for classifying agent log structuring approaches. Accordingly, existing approaches are organized into three categories: \textbf{Event and Schema Normalization}, \textbf{Semantic Artifact Construction}, and \textbf{Execution-Relation Construction} (see Fig.~\ref{fig:agent-log-structuring}). Table~\ref{tab:structuring-comparison} compares these approaches using the Structuring Design and Structuring Ability criteria introduced in Section~\ref{sec:structuring_criteria}. The following subsections review the three categories according to their resulting structured forms.

\begin{table*}[t]
\centering
\caption{Comparison of Agent Log Structuring Studies}
\label{tab:structuring-comparison}

\begingroup
\renewcommand{\arraystretch}{1.15}
\setlength{\tabcolsep}{4pt}
\scriptsize

\resizebox{0.95\textwidth}{!}{%
\begin{tabular}{
|>{\centering\arraybackslash}m{4.2cm}|
>{\centering\arraybackslash}m{3.5cm}|
>{\centering\arraybackslash}m{1.4cm}|
>{\centering\arraybackslash}m{1.2cm}|
>{\centering\arraybackslash}m{1.2cm}|
>{\centering\arraybackslash}m{1.2cm}|
>{\centering\arraybackslash}m{1.2cm}|
}

\hline


\multirow{2}{*}{\textbf{Study}}
&
\multicolumn{3}{c|}{\textbf{Structuring Design}}
&
\multicolumn{3}{c|}{\textbf{Structuring Ability}}
\\
\cline{2-7}

&
\textbf{SM}
&
\textbf{OM}
&
\textbf{EK}
&
\textbf{MA}
&
\textbf{CT}
&
\textbf{HR}
\\
\hline


\multicolumn{7}{|c|}{\cellcolor{gray!25}
\textbf{\textcolor{sectionblue}{\ref{sec:event-schema-normalization}} Event and Schema Normalization}
}
\\
\hline

Song et al.~\cite{song2026adp}
&
Rule-based
&
Offline
&
N
&
N
&
Y
&
Y
\\
\hline

Zhang et al.~\cite{zhang2026vcc}
&
Rule-based
&
Offline
&
N
&
N
&
N.A.
&
Y
\\
\hline

Sharma et al.~\cite{sharma2026willful}
&
Rule-based
&
Offline
&
N
&
N
&
Y
&
Y
\\
\hline

Barke et al.~\cite{barke2026agentrx}
&
Rule-based
&
Offline
&
N
&
Y
&
Y
&
Y
\\
\hline

Billiris and Gill~\cite{billiris2026foap}
&
LLM-based
&
Online
&
N
&
Y
&
Y
&
Y
\\
\hline


\multicolumn{7}{|c|}{\cellcolor{gray!25}
\textbf{\textcolor{sectionblue}{\ref{sec:semantic-artifact-construction}} Semantic Artifact Construction}
}
\\
\hline

Sriraman et al.~\cite{sriraman2026contextviewer}
&
LLM-based
&
Offline
&
Y
&
N
&
Y
&
N
\\
\hline

Rombaut et al.~\cite{rombaut2025watson}
&
LLM-based
&
Offline
&
N
&
N
&
Y
&
N
\\
\hline

Desmond et al.~\cite{desmond2025trajectoryexplorer}
&
Rule-based
&
Offline
&
N
&
N
&
N.A.
&
Y
\\
\hline

Zhuang et al.~\cite{zhuang2026trajectories}
&
LLM-based
&
Offline
&
Y
&
N
&
Y
&
Y
\\
\hline

Liang et al.~\cite{liang2026mcma}
&
Learning-based
&
Offline
&
N
&
N
&
Y
&
N
\\
\hline

Sheng et al.~\cite{sheng2026dills}
&
LLM-based
&
Offline
&
Y
&
Y
&
N
&
Y
\\
\hline

\multicolumn{7}{|c|}{\cellcolor{gray!25}
\textbf{\textcolor{sectionblue}{\ref{sec:execution-relation-construction}} Execution-Relation Construction}
}
\\
\hline

Souza et al.~\cite{souza2025provagent}
&
Rule-based
&
Online
&
Y
&
N
&
N.A.
&
Y
\\
\hline

Gao et al.~\cite{gao2026graphtrace}
&
LLM-based
&
Online
&
Y
&
N
&
Y
&
N
\\
\hline

Liu et al.~\cite{liu2026process}
&
Rule-based
&
Online
&
Y
&
N
&
Y
&
Y
\\
\hline

Fournier et al.~\cite{fournier2025processobservability}
&
Learning-based
&
Offline
&
N
&
Y
&
N.A.
&
N
\\
\hline

Fan et al.~\cite{fan2025metaflow}
&
Learning-based
&
Offline
&
Y
&
N
&
Y
&
N
\\
\hline

Qian et al.~\cite{qian2026memobrain}
&
Learning-based
&
Online
&
N
&
N
&
Y
&
Y
\\
\hline

Epperson et al.~\cite{epperson2025interactive}
&
Human-involved
&
Online
&
Y
&
Y
&
N.A.
&
N
\\
\hline

Lu et al.~\cite{lu2025agentlens}
&
LLM-based
&
Online
&
Y
&
Y
&
N.A.
&
Y
\\
\hline

Li et al.~\cite{li2026delinearizing}
&
Learning-based
&
Offline
&
N
&
N
&
Y
&
N
\\
\hline

Wu et al.~\cite{wu2026agentgraph}
&
LLM-based
&
Offline
&
N
&
N.A.
&
N.A.
&
Y
\\
\hline

Wang et al.~\cite{wang2026chief}
&
LLM-based
&
Offline
&
Y
&
Y
&
N.A.
&
Y
\\
\hline

Rafi et al.~\cite{rafi2026falat}
&
LLM-based
&
Offline
&
Y
&
Y
&
Y
&
Y
\\
\hline

Bakish et al.~\cite{bakish2026aig}
&
LLM-based
&
Offline
&
N
&
Y
&
N.A.
&
Y
\\
\hline

Wang~\cite{wang2026agenttracerca}
&
Rule-based
&
Offline
&
N
&
Y
&
Y
&
Y
\\
\hline

Cho et al.~\cite{cho2026automata}
&
Rule-based
&
Offline
&
N
&
Y
&
Y
&
Y
\\
\hline

Gueroudji et al.~\cite{gueroudji2026provenance}
&
Rule-based
&
Online
&
Y
&
Y
&
Y
&
Y
\\
\hline


\multicolumn{7}{|p{0.95\textwidth}|}{
\scriptsize
Structuring Design: design characteristics of the structuring approach;
Structuring Ability: applicability and capability of the structuring approach;
Y: Yes; N: No; N.A.: Not Available or Unknown;
SM: Structuring Mechanism, including Rule-based, LLM-based, Human-involved, and Learning-based;
OM: Operation Mode;
EK: External Knowledge Utilization;
MA: Multi-Agent Structuring;
CT: Cross-Log Transferability;
HR: Heterogeneous Record Structuring.
}
\\
\hline

\end{tabular}%
}

\endgroup
\end{table*}

\subsubsection{Event and Schema Normalization}
\label{sec:event-schema-normalization}

Agent execution records produced by different frameworks, datasets, or execution environments often differ in action spaces, observation formats, message structures, and trace schemas. To handle such heterogeneity, existing approaches introduce standardized schemas or intermediate representations to reorganize recorded execution information (e.g., ADP~\cite{song2026adp} and AgentRx~\cite{barke2026agentrx}).

A common strategy is to explicitly map source-specific records into a predefined schema or intermediate representation. ADP~\cite{song2026adp} converts dataset-specific actions and observations into standardized action--observation types, while AgentPex~\cite{sharma2026willful} uses modular adapters to normalize heterogeneous traces into a format-agnostic artifact containing prompts, tool schemas, message histories, and metadata. AgentRx~\cite{barke2026agentrx} similarly converts heterogeneous execution trajectories into a common intermediate representation for subsequent analysis. These approaches provide explicit correspondences between source records and normalized fields, but require source-specific mappings or adapters when unsupported formats are introduced~\cite{song2026adp,sharma2026willful}.

Rather than directly mapping records to a predefined schema, VCC~\cite{zhang2026vcc} adopts a compiler-style \textit{lex--parse--IR--lower--emit} pipeline. It first converts raw traces into a typed intermediate representation and then derives multiple structure-preserving views from the same representation (for more details, refer to Reference~\cite{zhang2026vcc}). This design separates source parsing from downstream trace organization and allows different views to share a common structural basis. However, source-specific lexers and parsers are still required for new trace formats.

FOAP~\cite{billiris2026foap} instead moves more adaptation into the normalization layer itself. It combines schema-agnostic telemetry ingestion with LLM-assisted normalization to construct canonical event records during execution while retaining the corresponding raw events. Compared with deterministic mappings and parsers, this design is more tolerant of changing event structures, but its prototype evaluation shows that greater event heterogeneity can reduce normalization success and that higher concurrency and event complexity increase processing latency~\cite{billiris2026foap}.

\begin{tcolorbox}[
    colback=summarybg,
    colframe=black,
    boxrule=0.5pt,
    arc=0pt,
    left=5pt,
    right=5pt,
    top=4pt,
    bottom=4pt
]
\textbf{Summary.}
Overall, existing approaches differ mainly in where source heterogeneity is resolved during structuring. Most adopt \textbf{rule-based and offline normalization} through source-specific mappings, adapters, or parsers (e.g., ADP~\cite{song2026adp}, AgentPex~\cite{sharma2026willful}, AgentRx~\cite{barke2026agentrx}, and VCC~\cite{zhang2026vcc}), whereas FOAP~\cite{billiris2026foap} provides an \textbf{LLM-based and online alternative}. This distinction reflects whether source variation is handled before normalization through explicit adaptation or absorbed within the normalization process itself. However, achieving \textbf{structural uniformity} does not necessarily ensure \textbf{semantic equivalence} across different sources (see Finding~5 in Section~4.3). In summary, these approaches establish a common structural basis for heterogeneous agent logs, while broader reuse still depends on preserving semantic consistency across normalized records.
\end{tcolorbox}

\subsubsection{Semantic Artifact Construction}
\label{sec:semantic-artifact-construction}

Structural normalization alone does not make the semantic organization of execution records explicit (see Section~\ref{sec:event-schema-normalization}). Semantic Artifact Construction therefore derives higher-level segments, annotations, summaries, or reconstructed reasoning from recorded content.

Some approaches organize existing content into source-grounded semantic units. Context Viewer~\cite{sriraman2026contextviewer} uses model-assisted processing to decompose LLM contexts into semantic artifacts, while Agent Trajectory Explorer~\cite{desmond2025trajectoryexplorer} supports visual inspection and manual trajectory annotation. DiLLS~\cite{sheng2026dills} summarizes multi-agent execution at multiple levels, covering overall planning and execution, action arrangements and outcomes, and concrete operation logs. These approaches retain relatively direct correspondence with source execution, but their semantic organization may depend on model interpretation, analyst judgement, or summarization granularity.

Other approaches construct information absent from the original record. Watson~\cite{rombaut2025watson} reconstructs candidate reasoning artifacts from observable agent inputs and outputs. Unlike segmentation or annotation, this introduces reasoning inferred rather than emitted by the original agent, increasing the distance between constructed artifacts and directly observed evidence.

Semantic construction can also aggregate multiple observations. From Trajectories to Evidence~\cite{zhuang2026trajectories} consolidates distributed observations into higher-level claim records with supporting evidence, provenance, and applicability boundaries. Compared with trajectory segmentation or reasoning reconstruction, explicitly linking claims to their evidential support places stronger constraints on which claims are retained.

\begin{tcolorbox}[
    colback=summarybg,
    colframe=black,
    boxrule=0.5pt,
    arc=0pt,
    left=5pt,
    right=5pt,
    top=4pt,
    bottom=4pt
]
\textbf{Summary.}
Compared with Event and Schema Normalization, which focuses on structural consistency (see Section~\ref{sec:event-schema-normalization}), Semantic Artifact Construction makes weakly expressed execution meaning explicit. Existing approaches mainly differ in their \textbf{distance from recorded evidence}: source-grounded organization retains closer correspondence with the original trajectory~\cite{sriraman2026contextviewer,desmond2025trajectoryexplorer,sheng2026dills}, whereas reconstructed or aggregated artifacts introduce greater semantic abstraction~\cite{rombaut2025watson,zhuang2026trajectories}. However, richer semantic organization does not ensure \textbf{semantic stability}, since the same execution evidence may yield different boundaries, labels, or reconstructed reasoning structures~\cite{sriraman2026contextviewer,desmond2025trajectoryexplorer,rombaut2025watson} (see Finding~6 in Section~4.3). In summary, these approaches improve the semantic interpretability of agent logs, while their broader reuse depends on maintaining both evidence grounding and consistent semantic organization.
\end{tcolorbox}

\subsubsection{Execution-Relation Construction}
\label{sec:execution-relation-construction}

Beyond normalization and semantic organization (see Sections~\ref{sec:event-schema-normalization} and~\ref{sec:semantic-artifact-construction}), Execution-Relation Construction makes connections among agents, tasks, events, and states explicit.

Some approaches map recorded or reported execution structure into relations. PROV-AGENT~\cite{souza2025provagent} constructs provenance lineage, Graph of Trace~\cite{gao2026graphtrace} organizes agent-reported subtask dependencies, and Graphectory/Langutory~\cite{liu2026process} maps actions to predefined process stages. Gueroudji et al.~\cite{gueroudji2026provenance} link agent activities to downstream workflow artifacts across heterogeneous runtimes. These structures retain relatively clear source links but depend on record completeness, reported dependencies, or predefined mappings.

Other approaches discover implicit or recurring structure from traces. AgentLens~\cite{lu2025agentlens} derives behavioral hierarchies from runtime events, whereas AgentGraph~\cite{wu2026agentgraph} extracts typed entities and relations linked to supporting trace spans. BPOP~\cite{li2026delinearizing} infers partial-order constraints from ordering variations across repeated trajectories, while process discovery~\cite{fournier2025processobservability} and finite-state abstraction~\cite{cho2026automata} derive recurring behavioral models and shared behavioral topologies, respectively. These approaches reveal structure beyond explicitly recorded relations but require additional abstraction or comparable traces.

For failure analysis, CHIEF~\cite{wang2026chief} constructs hierarchical causal graphs, while FALAT~\cite{rafi2026falat} traces dependencies to distinguish error-introducing steps from propagated failures. AIG~\cite{bakish2026aig} builds adaptive influence graphs for navigating failed trajectories, whereas AgentTrace~\cite{wang2026agenttracerca} reconstructs causal graphs for backward tracing from errors. Model-assisted construction can capture richer semantic dependencies, whereas source-linked or structurally derived relations offer more explicit grounding in recorded evidence. AGDebugger~\cite{epperson2025interactive} uses recoverable checkpoints and modified interactions to create alternative multi-agent execution branches. Unlike provenance or inferred dependencies, these branches represent alternative continuations rather than the original execution history.

\begin{tcolorbox}[
    colback=summarybg,
    colframe=black,
    boxrule=0.5pt,
    arc=0pt,
    left=5pt,
    right=5pt,
    top=4pt,
    bottom=4pt
]
\textbf{Summary.}
Execution-Relation Construction makes dependencies among execution elements explicit, whereas Event and Schema Normalization provides structural consistency and Semantic Artifact Construction makes execution meaning explicit (see Sections~\ref{sec:event-schema-normalization} and~\ref{sec:semantic-artifact-construction}, respectively).
Approaches differ mainly in the \textbf{evidential basis of relations}: preserving or mapping recorded or agent-reported relations retains closer grounding in recorded evidence~\cite{souza2025provagent,gao2026graphtrace}; deriving dependencies from execution traces can reveal implicit relations~\cite{li2026delinearizing,wang2026chief,rafi2026falat}; and interactive debugging constructs alternative execution branches through modified interactions~\cite{epperson2025interactive}.
However, stronger \textbf{relational explicitness} does not ensure equally explicit organization of \textbf{local execution context} (see Finding~7 in Section~4.3).
Broader use depends on preserving traceable links between constructed relations and their supporting evidence.
\end{tcolorbox}

\subsection{Discussion}
\label{sec:structuring-discussion}

\textbf{Finding 4: Cross-Level Structural Coherence.}
\textbf{Existing structuring approaches provide complementary representations at different abstraction levels, but explicit correspondence across these levels remains limited.}
Event and Schema Normalization preserves relatively fine-grained execution records, Semantic Artifact Construction organizes them into higher-level semantic units, and Execution-Relation Construction makes dependencies among execution elements explicit (see Sections~\ref{sec:event-schema-normalization}--\ref{sec:execution-relation-construction}). For example, VCC preserves links between structured views and its canonical trace~\cite{zhang2026vcc}, while AgentGraph associates extracted graph elements with supporting trace spans~\cite{wu2026agentgraph}. However, such source links mainly establish traceability within a representation rather than systematic mappings among event-level records, semantic artifacts, and execution relations. Consequently, the same execution element may appear at several abstraction levels without an explicit representation of how these views correspond. Current approaches therefore provide strong organization within individual levels but still lack \textbf{cross-level structural coherence}.

\textbf{Finding 5: Structural Uniformity and Semantic Equivalence.}
\textbf{Structurally normalized execution records are not necessarily semantically equivalent across heterogeneous sources.}
Existing normalization approaches map source-specific records into shared or typed representations, such as ADP's common action--observation representation~\cite{song2026adp} and VCC's typed intermediate representation~\cite{zhang2026vcc}. These mechanisms improve structural consistency and make heterogeneous traces easier to process under a common form. However, their normalized elements still depend on source-specific mappings, parsers, and structural conventions. When a new source introduces different action semantics or trace conventions, these adaptation rules must be extended accordingly~\cite{song2026adp,zhang2026vcc}. Thus, assigning records from different sources to the same structural type does not by itself establish that they represent the same execution semantics. Existing approaches therefore achieve stronger \textbf{structural uniformity} than \textbf{semantic equivalence}, which limits reliable reuse of the same interpretation across heterogeneous agent logs.

\textbf{Finding 6: Semantic Structure Stability.}
\textbf{Making execution semantics explicit does not guarantee that the resulting semantic organization is stable across different construction processes.}
Semantic Artifact Construction derives higher-level units through model-assisted segmentation, human annotation, reasoning reconstruction, or evidence aggregation~\cite{sriraman2026contextviewer,desmond2025trajectoryexplorer,rombaut2025watson,zhuang2026trajectories}. These mechanisms improve the interpretability of weakly structured execution content, but they also introduce construction choices that are not determined solely by the original trace. Model-based decomposition depends on the analysis configuration, human annotation depends on analyst judgement, and reconstructed reasoning introduces inferred content that was not directly emitted during execution~\cite{sriraman2026contextviewer,desmond2025trajectoryexplorer,rombaut2025watson}. Evidence grounding can constrain which higher-level artifacts are retained~\cite{zhuang2026trajectories}; however, traceability to source evidence does not by itself ensure that repeated structuring will yield the same semantic boundaries or organization. Consequently, current approaches improve semantic interpretability without yet establishing a common basis for \textbf{semantic stability}.

\textbf{Finding 7: Relational Explicitness and Local Context Granularity.}
\textbf{Execution-relation structures make dependencies more explicit, while local execution context is often represented at a different structural granularity.}
Existing approaches expose lineage, workflow, process, behavioral, and dependency relations through graph- or relation-oriented representations (e.g., PROV-AGENT~\cite{souza2025provagent}, Graph of Trace~\cite{gao2026graphtrace}, AgentGraph~\cite{wu2026agentgraph}, and BPOP~\cite{li2026delinearizing}). These structures substantially improve the visibility of how execution elements are connected. However, detailed information associated with individual elements, such as messages, arguments, observations, and runtime states, is commonly retained as node attributes, summaries, attached evidence, or information recoverable from the original trace rather than organized at an equivalent relational granularity. AgentGraph links graph elements back to trace spans~\cite{wu2026agentgraph}, while AGDebugger retains the messages and states underlying alternative execution branches~\cite{epperson2025interactive}; such mechanisms improve recoverability but do not remove the granularity difference between global relations and local context. Thus, current approaches make execution relations explicit but often leave the supporting local context less structured. The correspondence between higher-level relations and their detailed supporting evidence therefore remains unclear.

\section{Agent Log Mining}
\label{sec:agent-log-mining}

Agent log mining is the process of analyzing agent execution records to extract behavioral patterns and analytical information about agent execution. Unlike agent logging and agent log structuring, which focus on generating and organizing execution records, agent log mining uses the recorded execution evidence to identify abnormal execution behavior, diagnose failures, assess execution processes, and evaluate trustworthiness-related concerns. These analyses support the understanding and evaluation of agent execution behavior and provide analytical results for subsequent diagnosis, assessment, and system improvement.

\begin{figure*}[t]
\centering
\includegraphics[width=\textwidth]{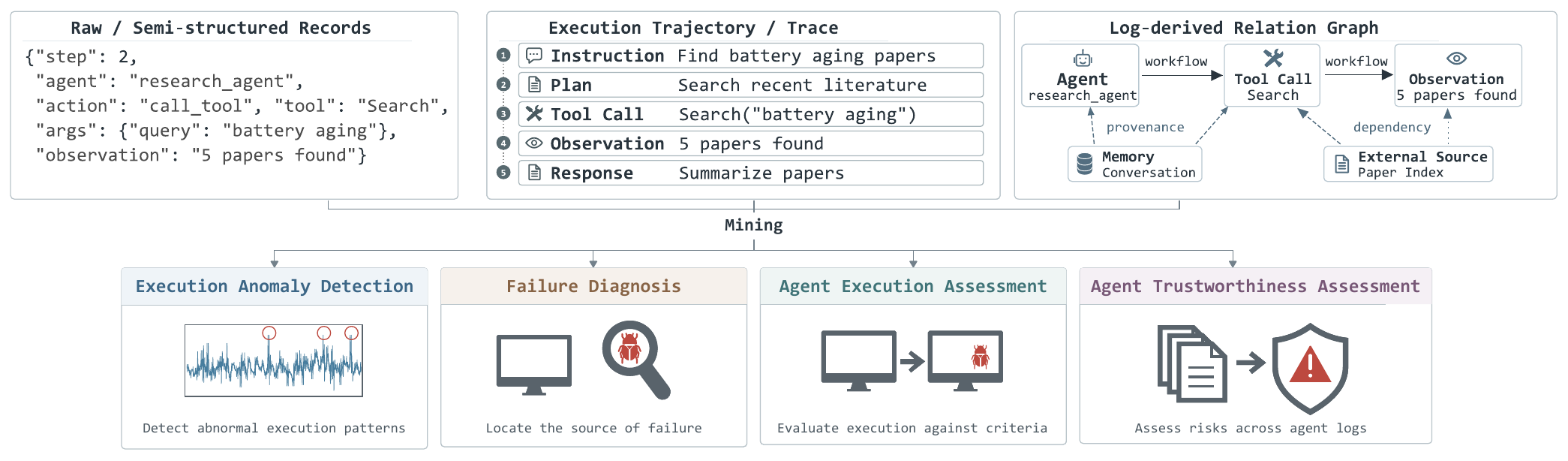}
\caption{Categories of agent log mining.}
\label{fig:agent-log-mining-overview}
\end{figure*}

\subsection{Comparison Criteria for Agent Log Mining Approaches}
\label{sec:mining-evaluation-criteria}

In this section, we propose a uniform set of comparison criteria for reviewing and comparing different agent log mining approaches (see Table~\ref{tab:mining-criteria}). This set of criteria characterizes agent log mining approaches in two dimensions: Mining Design and Mining Ability. These two dimensions contain six metrics, which reflect the similarities and differences among existing agent log mining approaches from different perspectives.

\begin{table*}[t]
\centering
\caption{Comparison Dimensions and Metrics for Agent Log Mining Approaches}
\label{tab:mining-criteria}

\begingroup
\renewcommand{\arraystretch}{1.12}
\setlength{\tabcolsep}{4pt}
\scriptsize

\begin{tabularx}{\textwidth}{
|>{\raggedright\arraybackslash}p{0.275\textwidth}|
>{\raggedright\arraybackslash}X|
}
\hline

\centering\textbf{Dimension} &
\centering\arraybackslash\textbf{Metrics}
\\
\hline

\textbf{Mining Design.}
This dimension describes how an approach organizes and analyzes agent
execution evidence.
Different approaches vary in the execution context they consider, the
principal mining mechanism they employ, and their dependence on external
knowledge or references.
Overall, we use three metrics to characterize mining design.
&
\textbf{(1) Execution-Context Scope (EC).}
This metric indicates the principal scope of execution evidence considered
by the mining approach.
We distinguish local context (L), trajectory-level context (T), and
relational context (R), corresponding respectively to bounded execution
evidence, complete or extended trajectories, and explicit relations among
execution entities.

\par\smallskip\hrule\smallskip

\textbf{(2) Mining Mechanism (MM).}
This metric indicates the principal mechanism used to derive mining results
from agent execution evidence.
We distinguish statistical or pattern-based analysis (SP), model-based
analysis (MB), rule or structural analysis (RS), and hybrid analysis (HY).

\par\smallskip\hrule\smallskip

\textbf{(3) Knowledge and Reference Basis (KR).}
This metric indicates whether the mining process relies on explicit
knowledge or reference information beyond the execution evidence being
analyzed.
Such information may include task references, diagnostic knowledge,
policies, or evaluation criteria.
\\
\hline

\textbf{Mining Ability.}
This dimension describes the applicability and operational capability of
agent log mining approaches.
Different approaches vary in their support for multi-agent executions,
their generalizability across agent systems, and whether they can operate
during execution.
We use three metrics to characterize these differences.
&
\textbf{(4) Multi-Agent Applicability (MA).}
This metric indicates whether the approach is designed or demonstrated to
analyze execution logs from multi-agent systems.
Analysis of multiple independent agents or models alone does not establish
multi-agent applicability.

\par\smallskip\hrule\smallskip

\textbf{(5) Agent Generalizability (AG).}
This metric indicates whether the same core mining approach is demonstrated
to work across different agent systems or architectures.
Evaluation on different datasets, tasks, or model backbones alone does not
establish such generalizability.

\par\smallskip\hrule\smallskip

\textbf{(6) Online Analysis (OA).}
This metric indicates whether the approach can analyze incrementally
generated execution evidence and produce its principal result while the
corresponding agent execution is still ongoing.
Approaches that require completed execution records are considered offline.
\\
\hline

\end{tabularx}

\endgroup
\end{table*}

\subsection{Taxonomy of Agent Log Mining Techniques}
\label{sec:mining-taxonomy}

We use the primary analytical objective as the basis for classifying agent log mining techniques. Accordingly, existing studies are organized into four categories: \textbf{Execution Anomaly Detection}, \textbf{Failure Diagnosis}, \textbf{Agent Execution Assessment}, and \textbf{Agent Trustworthiness Assessment} (see Fig.~\ref{fig:agent-log-mining-overview}). Table~\ref{tab:mining-comparison} compares these studies using the \textbf{Mining Design} and \textbf{Mining Ability} criteria introduced in Section~\ref{sec:mining-evaluation-criteria}. The following subsections review the four categories according to their primary analytical objectives.

\begin{table*}[t]
\centering
\caption{Comparison of Agent Log Mining Studies}
\label{tab:mining-comparison}

\begingroup
\renewcommand{\arraystretch}{1.08}
\setlength{\tabcolsep}{2.2pt}
\scriptsize

\begin{tabularx}{0.95\textwidth}{
|>{\centering\arraybackslash}p{0.22\textwidth}|
>{\centering\arraybackslash}p{0.0295\textwidth}|
>{\centering\arraybackslash}p{0.0328\textwidth}|
>{\centering\arraybackslash}p{0.0295\textwidth}|
>{\centering\arraybackslash}p{0.0295\textwidth}|
>{\centering\arraybackslash}p{0.0295\textwidth}|
>{\centering\arraybackslash}p{0.0295\textwidth}|
>{\centering\arraybackslash}p{0.22\textwidth}|
>{\centering\arraybackslash}p{0.0295\textwidth}|
>{\centering\arraybackslash}p{0.0328\textwidth}|
>{\centering\arraybackslash}p{0.0295\textwidth}|
>{\centering\arraybackslash}p{0.0295\textwidth}|
>{\centering\arraybackslash}p{0.0295\textwidth}|
>{\centering\arraybackslash}p{0.0295\textwidth}|
}
\hline

&
\multicolumn{3}{c|}{\textbf{Mining Design}}
&
\multicolumn{3}{c|}{\textbf{Mining Ability}}
&

&
\multicolumn{3}{c|}{\textbf{Mining Design}}
&
\multicolumn{3}{c|}{\textbf{Mining Ability}}
\\

\cline{2-7}\cline{9-14}

\multirow{-2}{*}{\textbf{Study}}
&
\textbf{EC} &
\textbf{MM} &
\textbf{KR} &
\textbf{MA} &
\textbf{AG} &
\textbf{OA}
&

\multirow{-2}{*}{\textbf{Study}}
&
\textbf{EC} &
\textbf{MM} &
\textbf{KR} &
\textbf{MA} &
\textbf{AG} &
\textbf{OA}
\\
\hline


\multicolumn{14}{|c|}{\rule{0pt}{1.1ex}}
\\
\hline


\multicolumn{7}{|c|}{
\cellcolor{gray!25}
\textbf{\textcolor{sectionblue}{\ref{sec:execution-anomaly-detection}} Execution Anomaly Detection}
}
&
Luan et al.~\cite{luan2026repair}
& R & HY & Y & Y & Y & N
\\
\hline

\multicolumn{7}{|c|}{
\textbf{Execution-Signal Detection}
}
&
\multicolumn{7}{c|}{\rule{0pt}{1.1ex}}
\\
\hline

Pathak et al.~\cite{pathak2025detectingsilent}
& T & MB & Y & Y & Y & N
&
\multicolumn{7}{c|}{
\cellcolor{gray!25}
\textbf{\textcolor{sectionblue}{\ref{sec:agent-execution-assessment}} Agent Execution Assessment}
}
\\
\hline

Sanna~\cite{sanna2025cross}
& T & MB & Y & N.A. & N.A. & Y
&
\multicolumn{7}{c|}{
\textbf{Process Quality Evaluation}
}
\\
\hline

George et al.~\cite{george2026cycledetection}
& R & HY & N & Y & N & N
&
Ma et al.~\cite{ma2024agentboard}
& T & RS & Y & N.A. & Y & Y
\\
\hline

\multicolumn{7}{|c|}{
\textbf{Contextual Behavior Monitoring}
}
&
Zhang et al.~\cite{zhang2024llamatouch}
& T & RS & Y & N.A. & Y & N
\\
\hline

Storf et al.~\cite{storf2026constitutional}
& T & MB & Y & N.A. & Y & N
&
Gritta et al.~\cite{gritta2026process}
& T & MB & Y & Y & N.A. & N
\\
\hline

Kale et al.~\cite{kale2025reliable}
& T & MB & Y & N.A. & Y & N
&
Mallikarachchi~\cite{mallikarachchi2026autogentracekit}
& T & SP & N.A. & N.A. & N & N
\\
\hline

Arike et al.~\cite{arike2026does}
& T & MB & Y & N.A. & Y & N
&
Kim et al.~\cite{kim2025beyond}
& T & MB & Y & Y & Y & N
\\
\hline

Vaknin et al.~\cite{vaknin2026mastitch}
& R & MB & Y & Y & Y & Y
&
Chen et al.~\cite{chen2026traceevaluation}
& T & HY & Y & N.A. & Y & N
\\
\hline

Solomon et al.~\cite{solomon2026lumimas}
& T & MB & Y & Y & Y & Y
&
Yehudai et al.~\cite{yehudai2026agentic}
& R & MB & Y & Y & Y & N
\\
\hline

Mou et al.~\cite{mou2026toolsafe}
& T & MB & Y & N.A. & Y & Y
&
Ou et al.~\cite{ou2025agentdiagnose}
& T & MB & Y & N.A. & N & N
\\
\hline


\multicolumn{7}{|c|}{\rule{0pt}{1.1ex}}
&
Aghzal et al.~\cite{aghzal2026webagentsfail}
& T & HY & Y & N.A. & N.A. & Y
\\
\hline

\multicolumn{7}{|c|}{
\cellcolor{gray!25}
\textbf{\textcolor{sectionblue}{\ref{sec:failure-diagnosis}} Failure Diagnosis}
}
&
\multicolumn{7}{c|}{
\textbf{Execution Verification}
}
\\
\hline

\multicolumn{7}{|c|}{
\textbf{Failure Attribution}
}
&
Patel~\cite{patel2026runtimeverification}
& R & HY & Y & Y & N.A. & Y
\\
\hline

Zhang et al.~\cite{zhang2025agentcauses}
& T & MB & Y & Y & Y & N
&
Zheng et al.~\cite{zheng2025audagent}
& R & RS & Y & N & Y & Y
\\
\hline

Zhu et al.~\cite{zhu2026raffles}
& T & HY & Y & Y & Y & N
&
Alvarez et al.~\cite{alvarez2026provenanceguard}
& R & HY & Y & N & N & N
\\
\hline

Sun et al.~\cite{sun2026scope}
& T & MB & Y & Y & Y & N
&
Grigor et al.~\cite{grigor2026vet}
& R & RS & Y & N & N.A. & N
\\
\hline

Geng et al.~\cite{geng2026failurelocalization}
& R & MB & Y & Y & Y & N
&
\multicolumn{7}{c|}{
\cellcolor{gray!25}
\textbf{\textcolor{sectionblue}{\ref{sec:agent-trustworthiness-assessment}} Agent Trustworthiness Assessment}
}
\\
\hline

Zhang et al.~\cite{zhang2025agentracer}
& T & MB & Y & Y & Y & N
&
\multicolumn{7}{c|}{
\textbf{Security Risk Assessment}
}
\\
\hline

Kong et al.~\cite{kong2026aegis}
& T & MB & Y & Y & Y & N
&
Ruan et al.~\cite{ruan2024toolemu}
& T & MB & Y & N.A. & N.A. & N
\\
\hline

Qiao et al.~\cite{qiao2026verifymas}
& T & MB & Y & Y & Y & N
&
Zhang et al.~\cite{zhang2025agentsecuritybench}
& L & RS & Y & N.A. & Y & N
\\
\hline

Ge et al.~\cite{ge2026spectrum}
& T & SP & Y & Y & Y & N
&
Ying et al.~\cite{ying2025securewebarena}
& T & HY & Y & N.A. & Y & N
\\
\hline

Li et al.~\cite{li2026self}
& R & HY & Y & Y & Y & N
&
Zhang et al.~\cite{zhang2025mcp}
& L & RS & Y & N.A. & N.A. & N
\\
\hline

Yu et al.~\cite{yu2026correct}
& T & MB & Y & Y & Y & N
&
Ersoy et al.~\cite{ersoy2026darkpatterns}
& T & RS & Y & N.A. & Y & N
\\
\hline

Ma et al.~\cite{ma2026dover}
& T & MB & Y & Y & Y & N
&
Atinafu et al.~\cite{atinafu2026rewardhacking}
& L & RS & Y & N & N.A. & N
\\
\hline

Zhu et al.~\cite{zhu2026stepfinder}
& T & MB & Y & Y & Y & N
&
\multicolumn{7}{c|}{
\textbf{Privacy Leakage Assessment}
}
\\
\hline

\multicolumn{7}{|c|}{
\textbf{Diagnostic Support}
}
&
Biswas et al.~\cite{biswas2025privacy}
& R & SP & Y & Y & N.A. & N.A.
\\
\hline

Mulian et al.~\cite{mulian2026agentfixer}
& T & MB & Y & Y & N & N
&
El Yagoubi et al.~\cite{yagoubi2026agentleak}
& R & HY & Y & Y & Y & N
\\
\hline

Atreja~\cite{atreja2026pathfinder}
& R & HY & Y & Y & Y & N
&
Choong et al.~\cite{choong2026capchain}
& R & RS & Y & Y & N & Y
\\
\hline

Wang et al.~\cite{wang2026xagen}
& R & HY & Y & Y & N & N
&
\multicolumn{7}{c|}{}
\\
\hline


\multicolumn{14}{|c|}{\rule{0pt}{1.1ex}}
\\
\hline


\multicolumn{14}{|p{0.92\textwidth}|}{
\footnotesize
Mining Design: design characteristics of the mining approach;
Mining Ability: capability of the mining approach;
Y: Yes; N: No; N.A.: Not Available or Unknown;
EC: Execution-Context Scope;
MM: Mining Mechanism;
KR: Knowledge and Reference Basis;
MA: Multi-Agent Applicability;
AG: Agent Generalizability;
OA: Online Analysis;
L: Local context;
T: Trajectory-level context;
R: Relational context;
SP: Statistical Analysis;
MB: Model-based Analysis;
RS: Rule Analysis;
HY: Hybrid Analysis.
}
\\
\hline

\end{tabularx}

\endgroup
\end{table*}

\begin{table*}[t]
\centering
\caption{Task-Specific Characteristics of Execution Anomaly Detection Studies}
\label{tab:ead-characteristics}

\begingroup
\renewcommand{\arraystretch}{1.15}
\setlength{\tabcolsep}{4pt}
\scriptsize

\begin{tabularx}{0.95\textwidth}{
|>{\centering\arraybackslash}p{0.165\textwidth}|
>{\centering\arraybackslash}p{0.165\textwidth}|
>{\centering\arraybackslash}p{0.155\textwidth}|
>{\centering\arraybackslash}p{0.165\textwidth}|
>{\centering\arraybackslash}X|
}
\hline

\rowcolor{gray!20}
\textbf{Study} &
\textbf{Anomaly Target} &
\textbf{Detector Structure} &
\textbf{Detector Adaptation} &
\textbf{Detection Resolution}
\\
\hline

Pathak et al.~\cite{pathak2025detectingsilent}
&
Execution Failure
&
Single
&
Data
&
Execution
\\
\hline

Sanna~\cite{sanna2025cross}
&
Safety Risk
&
Single
&
Data
&
Execution
\\
\hline

George et al.~\cite{george2026cycledetection}
&
Execution Failure
&
Single
&
None
&
Execution
\\
\hline

Storf et al.~\cite{storf2026constitutional}
&
Safety Risk
&
Single
&
Data
&
Execution
\\
\hline

Kale et al.~\cite{kale2025reliable}
&
Safety Risk
&
Multi
&
None
&
Execution
\\
\hline

Arike et al.~\cite{arike2026does}
&
Safety Risk
&
Multi
&
None
&
Execution
\\
\hline

Vaknin et al.~\cite{vaknin2026mastitch}
&
Mixed
&
Multi
&
None
&
Localized
\\
\hline

Solomon et al.~\cite{solomon2026lumimas}
&
Mixed
&
Single
&
Data
&
Localized
\\
\hline

Mou et al.~\cite{mou2026toolsafe}
&
Safety Risk
&
Single
&
Data
&
Localized
\\
\hline

\multicolumn{5}{|p{0.92\textwidth}|}{
Anomaly Target: type of anomaly being detected;
Detector Structure: organization of the detection process;
Detector Adaptation: task-specific adaptation of the detector;
Detection Resolution: granularity of the detection result;
Execution Failure: execution-related anomalies;
Safety Risk: safety-related anomalies;
Mixed: both anomaly types;
Single: one detection stage;
Multi: multiple detection stages;
Data: adaptation using task-related data;
None: no task-specific adaptation;
Execution: overall-execution detection;
Localized: localization to specific execution elements.
}
\\
\hline

\end{tabularx}
\endgroup
\end{table*}

\subsubsection{Execution Anomaly Detection}
\label{sec:execution-anomaly-detection}

Execution Anomaly Detection identifies departures from normal or expected agent behavior, rather than localizing or explaining observed failures (see Section~\ref{sec:failure-diagnosis}). We distinguish \textbf{Execution-Signal Detection}, which detects anomalies using explicit behavioral, structural, or runtime signals, from \textbf{Contextual Behavior Monitoring}, which interprets richer trajectory context against task goals, behavioral expectations, or safety requirements. Table~\ref{tab:ead-characteristics} summarizes their task-specific characteristics.

\textbf{Execution-Signal Detection.}
Execution-Signal Detection identifies anomalies using explicit signals derived from trajectories rather than directly analyzing the original execution content. One pattern maps heterogeneous records into a comparable feature space and learns anomaly criteria from prior runs. Existing studies use different combinations of contextual, structural, temporal, sequential, action, data-flow, and system-level features to detect execution failures or behavioral backdoors~\cite{pathak2025detectingsilent,sanna2025cross}. Their differences lie mainly in the retained signals rather than the overall analytical process.

A second pattern tests explicit regularities in the current execution instead of learning normality from historical data. George et al.~\cite{george2026cycledetection} combine temporal call-stack repetition with semantic similarity of repeated content to identify cycles associated with insufficient progress. Thus, the two patterns differ in whether anomaly criteria are learned from prior runs or specified through execution regularities, but both convert trajectory evidence into explicit signals before producing execution-level judgments.

\textbf{Contextual Behavior Monitoring.}
Some behaviors become anomalous only relative to task intent, behavioral specifications, or safety requirements. Contextual Behavior Monitoring therefore retains richer semantic context and uses LLM-based monitors or analytical agents for judgment. Existing approaches differ in how evidence enters the monitoring process and at what granularity anomalies are identified.

One group emphasizes \emph{context organization and selection}. Monitor adaptation uses task specifications and synthetic trajectories to adjust judgments of messages, tool calls, and outputs~\cite{storf2026constitutional}; trajectory scaffolding organizes long trajectories through full-context, hierarchical, sequential, or hybrid structures~\cite{kale2025reliable}; and evidence selection separates relevant-fragment extraction from final judgment~\cite{arike2026does}. These designs alter the judging mechanism, evidence organization, and evidence reaching the judge, respectively. Contextual monitoring therefore depends not only on log volume but also on how records are transformed into effective analytical context.

A second group extends whole-execution verdicts toward \emph{localized anomaly evidence}. Methods decompose multi-agent traces into agent-level units and combine local judgments with global context~\cite{vaknin2026mastitch}, integrate learned execution deviations with explanatory analysis~\cite{solomon2026lumimas}, or assess candidate tool invocations alongside user requests and interaction histories~\cite{mou2026toolsafe}. These methods seek to link irregularities to specific behaviors or actions, differing mainly in analytical granularity: agent-level decomposition captures local behavior and cross-agent context, whereas tool-level monitoring evaluates concrete actions during ongoing execution.

\begin{tcolorbox}[
    colback=summarybg,
    colframe=black,
    boxrule=0.5pt,
    arc=0pt,
    left=5pt,
    right=5pt,
    top=4pt,
    bottom=4pt
]
\textbf{Summary.}
Execution Anomaly Detection identifies departures from normal or expected agent behavior.
\textbf{Execution-Signal Detection} applies learned decision boundaries or predefined patterns to trajectory-derived behavioral or structural signals, mainly yielding execution-level judgments~\cite{pathak2025detectingsilent,sanna2025cross,george2026cycledetection}.
\textbf{Contextual Behavior Monitoring} instead interprets richer trajectory context against behavioral or safety expectations and can extend detection to individual agents or tool actions~\cite{storf2026constitutional,kale2025reliable,vaknin2026mastitch,mou2026toolsafe}.
The two families differ mainly in \emph{judgment basis} and \emph{detection resolution}; contextual judgments also depend on evidence organization.
These complementary detection mechanisms support subsequent failure analysis, whereas failure diagnosis further attributes or explains observed failures (see Section~\ref{sec:failure-diagnosis}).
\end{tcolorbox}

\begin{table*}[t]
\centering
\caption{Task-Specific Characteristics of Failure Diagnosis Studies}
\label{tab:fd-characteristics}

\begingroup
\renewcommand{\arraystretch}{1.15}
\setlength{\tabcolsep}{2.5pt}
\scriptsize

\begin{tabularx}{0.95\textwidth}{
|>{\centering\arraybackslash}p{0.145\textwidth}
|>{\centering\arraybackslash}p{0.155\textwidth}
|>{\centering\arraybackslash}p{0.155\textwidth}
|>{\centering\arraybackslash}p{0.155\textwidth}
|>{\centering\arraybackslash}X|
}
\hline

\rowcolor{gray!20}
\textbf{Study} &
\textbf{Diagnosis Target} &
\textbf{Diagnostic Structure} &
\textbf{Localization Strategy} &
\textbf{Candidate Handling}
\\
\hline

Zhang et al.~\cite{zhang2025agentcauses}
& Mixed
& Multi
& Search
& Iterative Refinement
\\
\hline

Zhu et al.~\cite{zhu2026raffles}
& Step
& Multi
& Search
& Iterative Refinement
\\
\hline

Sun et al.~\cite{sun2026scope}
& Mixed
& Multi
& Search
& Iterative Refinement
\\
\hline

Geng et al.~\cite{geng2026failurelocalization}
& Agent
& Multi
& Learning
& Candidate Ranking
\\
\hline

Zhang et al.~\cite{zhang2025agentracer}
& Mixed
& Single
& Learning
& Direct Attribution
\\
\hline

Kong et al.~\cite{kong2026aegis}
& Agent
& Single
& Learning
& Direct Attribution
\\
\hline

Qiao et al.~\cite{qiao2026verifymas}
& Agent
& Multi
& Learning
& Direct Attribution
\\
\hline

Ge et al.~\cite{ge2026spectrum}
& Step
& Single
& Search
& Candidate Ranking
\\
\hline

Li et al.~\cite{li2026self}
& Mixed
& Multi
& Experience
& Iterative Refinement
\\
\hline

Yu et al.~\cite{yu2026correct}
& Mixed
& Multi
& Experience
& Direct Attribution
\\
\hline

Ma et al.~\cite{ma2026dover}
& Mixed
& Multi
& Execution
& Iterative Refinement
\\
\hline

Zhu et al.~\cite{zhu2026stepfinder}
& Step
& Single
& Learning
& Candidate Ranking
\\
\hline

Mulian et al.~\cite{mulian2026agentfixer}
& Agent
& Multi
& Experience
& Direct Attribution
\\
\hline

Atreja~\cite{atreja2026pathfinder}
& Step
& Multi
& Execution
& Iterative Refinement
\\
\hline

Wang et al.~\cite{wang2026xagen}
& Mixed
& Multi
& Hybrid
& Direct Attribution
\\
\hline

Luan et al.~\cite{luan2026repair}
& Step
& Multi
& Execution
& Candidate Ranking
\\
\hline

\multicolumn{5}{
|>{\raggedright\arraybackslash}p{\dimexpr0.95\textwidth-2\tabcolsep-2\arrayrulewidth\relax}|
}{
Diagnosis Target: granularity of the diagnosed failure source;
Diagnostic Structure: organization of the diagnostic process;
Localization Strategy: strategy used to identify failure sources;
Candidate Handling: method used to process candidate failure sources;
Agent: agent-level localization;
Step: step-level localization;
Mixed: both localization levels;
Single: one diagnostic stage;
Multi: multiple diagnostic stages;
Search: candidate search or narrowing;
Learning: learned attribution;
Experience: historical diagnostic experience;
Execution: executable analysis or intervention;
Hybrid: combined automated and human-guided localization;
Direct Attribution: direct failure-source identification;
Candidate Ranking: ranking or scoring of potential failure sources;
Iterative Refinement: progressive refinement of failure candidates.
}
\\
\hline

\end{tabularx}
\endgroup
\end{table*}

\subsubsection{Failure Diagnosis}
\label{sec:failure-diagnosis}

A failed agent execution indicates that the intended task outcome has not been achieved, but the final failure alone does not reveal which execution element introduced the error, how it propagated, or what information is needed for correction. Unlike execution anomaly detection, which determines whether abnormal behavior is present, failure diagnosis starts from an observed failure and analyzes execution evidence to identify responsible agents or steps and, in some cases, support explanation or remediation. Existing approaches differ mainly in diagnostic depth: some attribute failures to responsible execution elements, whereas others further transform diagnostic evidence into information for understanding or corrective action. Accordingly, we organize existing approaches into \textbf{Failure Attribution} and \textbf{Diagnostic Support}. Table~\ref{tab:fd-characteristics} summarizes their task-specific characteristics.

\textbf{Failure Attribution.}
Failure Attribution associates observed failures with responsible agents, steps, or combinations of execution elements. Long multi-agent trajectories may contain many plausible candidates, while agent- and step-level responsibility need not be equivalent. Who\&When treats the responsible agent and decisive error step as separate attribution targets~\cite{zhang2025agentcauses}. Existing approaches therefore commonly narrow the attribution space before judgment, but differ in how they do so.

One pattern constrains hypotheses or candidate regions before localization. RAFFLES iteratively generates, evaluates, and refines fault hypotheses through a Judge--Evaluator loop~\cite{zhu2026raffles}; Scope Delineation identifies a likely failure scope before fine-grained localization~\cite{sun2026scope}; VerifyMAS checks predefined error hypotheses against the complete trajectory before locating responsible agents for supported hypotheses~\cite{qiao2026verifymas}. These methods seek to avoid unconstrained attribution, differing in \emph{what is constrained first}: competing explanations, execution scope, or failure type.

Another pattern uses learned representations or specialized predictors. FLKR estimates agent responsibility through behavioral representations, knowledge--strategy alignment, and consistency reasoning~\cite{geng2026failurelocalization}. AgenTracer and Aegis derive supervision through counterfactual replay or controlled fault injection to predict responsible agents, steps, or error modes~\cite{zhang2025agentracer,kong2026aegis}. StepFinder models steps as temporal semantic sequences and assigns step-level error scores using temporal and agent-aware dependencies~\cite{zhu2026stepfinder}. These methods shift from free-form per-trajectory reasoning to learned scoring or prediction, differing mainly in supervision sources and agent-, step-, or mixed-level attribution.

Attribution can also use evidence beyond the current failed trajectory. FAMAS computes spectrum-based suspiciousness scores for agents and actions across repeated runs of the same task~\cite{ge2026spectrum}. Experience-based methods instead reuse prior diagnoses: ERRORPROBE stores verified diagnostic episodes in episodic memory, whereas CORRECT compresses recurring failures into reusable error schemata~\cite{li2026self,yu2026correct}. The latter two retain historical evidence at different abstraction levels: case-specific episodes versus recurring failure structures.

Intervention provides a different evidential basis from correlation or historical reuse. DoVer treats a suspected source as a hypothesis, modifies the corresponding message, instruction, or plan, and re-executes the downstream workflow to test outcome changes~\cite{ma2026dover}. Thus, cross-execution statistical associations and historical diagnostic priors indicate plausibility, whereas controlled re-execution directly examines the effect of modifying the candidate itself.

\textbf{Diagnostic Support.}
Localization identifies where a failure originated or which agent was responsible, but it does not explain the failure, indicate how to correct it, or establish whether a proposed modification would resolve it. Diagnostic Support therefore transforms execution evidence and attribution results into explanations, recommendations, interactive debugging information, or repair-oriented analysis.

One pattern converts accumulated diagnostic evidence into corrective guidance. AgentFixer aggregates failed executions, validation results, and root-cause analyses to identify recurring failure patterns, prioritize problematic agents, and recommend modifications~\cite{mulian2026agentfixer}. XAgen combines an interactive workflow representation of logs with LLM-based error identification and human feedback, supporting failure inspection, configuration revision, and re-execution~\cite{wang2026xagen}. Both connect diagnosis to correction, but through automated remediation recommendations and iterative human-guided debugging, respectively.

A second pattern incorporates executable evidence. Pathfinder combines a learned trace analyzer with code execution to characterize case-specific failures and improves the analyzer through adversarial self-play~\cite{atreja2026pathfinder}. Repair or Resample uses SymTrace's event-dependency snapshots to replay execution up to an intervention boundary and apply localized modifications while keeping preceding execution fixed~\cite{luan2026repair}. Both go beyond textual explanation: Pathfinder strengthens failure characterization through executable analysis, whereas SymTrace tests localized corrections through controlled replay that reduces unrelated upstream variation.

\begin{tcolorbox}[
    colback=summarybg,
    colframe=black,
    boxrule=0.5pt,
    arc=0pt,
    left=5pt,
    right=5pt,
    top=4pt,
    bottom=4pt
]
\textbf{Summary.}
Failure Diagnosis analyzes failed executions to identify responsible execution elements and support subsequent understanding or correction. \textbf{Failure Attribution} localizes responsible agents or steps through structured hypothesis reasoning, learned attribution models, cross-execution or historical evidence, and intervention-based analysis~\cite{zhu2026raffles,qiao2026verifymas,geng2026failurelocalization,ge2026spectrum,ma2026dover}. \textbf{Diagnostic Support} extends diagnostic evidence toward explanations, remediation recommendations, interactive debugging, or executable repair analysis~\cite{mulian2026agentfixer,wang2026xagen,atreja2026pathfinder,luan2026repair}. Thus, the two families differ mainly in their \emph{diagnostic objective} and \emph{evidence depth}: the former focuses on establishing responsibility for an observed failure, whereas the latter transforms diagnostic evidence into information or actions that support correction. Failure diagnosis is closely related to anomaly detection, while their analytical objectives differ: anomaly detection determines whether abnormal behavior is present, whereas failure diagnosis localizes or explains an observed failure (see Section~\ref{sec:execution-anomaly-detection}). In summary, the key distinction lies not in which model performs diagnosis, but in how failed trajectories are transformed into increasingly actionable evidence about responsibility and corrective action.
\end{tcolorbox}

\begin{table*}[t]
\centering
\caption{Task-Specific Characteristics of Agent Execution Assessment Studies}
\label{tab:aea-characteristics}

\begingroup
\renewcommand{\arraystretch}{1.15}
\setlength{\tabcolsep}{3.5pt}
\scriptsize

\begin{tabularx}{0.95\textwidth}{
|>{\centering\arraybackslash}p{0.155\textwidth}
|>{\centering\arraybackslash}p{0.155\textwidth}
|>{\centering\arraybackslash}p{0.155\textwidth}
|>{\centering\arraybackslash}p{0.155\textwidth}
|>{\centering\arraybackslash}X|
}
\hline

\rowcolor{gray!20}
\textbf{Study} &
\textbf{Assessment Target} &
\textbf{Assessment Structure} &
\textbf{Assessment Strategy} &
\textbf{Assessment Claim}
\\
\hline

Ma et al.~\cite{ma2024agentboard}
& State
& Single
& Matching
& Task Progress
\\
\hline

Zhang et al.~\cite{zhang2024llamatouch}
& State
& Single
& Matching
& Task Progress
\\
\hline

Gritta et al.~\cite{gritta2026process}
& Behavior
& Single
& Judging
& Requirement Satisfaction
\\
\hline

Mallikarachchi~\cite{mallikarachchi2026autogentracekit}
& Behavior
& Single
& Measurement
& Execution Quality
\\
\hline

Kim et al.~\cite{kim2025beyond}
& Artifact
& Single
& Judging
& Execution Quality
\\
\hline

Chen et al.~\cite{chen2026traceevaluation}
& Artifact
& Multi
& Measurement
& Execution Quality
\\
\hline

Yehudai et al.~\cite{yehudai2026agentic}
& Behavior
& Single
& Judging
& Execution Quality
\\
\hline

Ou et al.~\cite{ou2025agentdiagnose}
& Artifact
& Single
& Judging
& Execution Quality
\\
\hline

Aghzal et al.~\cite{aghzal2026webagentsfail}
& Behavior
& Single
& Measurement
& Execution Quality
\\
\hline

Patel~\cite{patel2026runtimeverification}
& Behavior
& Single
& Measurement
& Requirement Satisfaction
\\
\hline

Zheng et al.~\cite{zheng2025audagent}
& Behavior
& Single
& Verification
& Requirement Satisfaction
\\
\hline

Alvarez et al.~\cite{alvarez2026provenanceguard}
& Artifact
& Multi
& Verification
& Requirement Satisfaction
\\
\hline

Grigor et al.~\cite{grigor2026vet}
& Artifact
& Single
& Verification
& Requirement Satisfaction
\\
\hline

\multicolumn{5}{
|>{\raggedright\arraybackslash}p{\dimexpr0.95\textwidth-2\tabcolsep-2\arrayrulewidth\relax}|
}{
Assessment Target: object of the assessment;
Assessment Structure: organization of the assessment process;
Assessment Strategy: strategy used to derive the assessment result;
Assessment Claim: type of conclusion produced by the assessment;
State: task states or subgoals;
Behavior: execution behaviors or interactions;
Artifact: execution traces or generated outputs;
Single: one assessment stage;
Multi: multiple assessment stages;
Matching: matching execution evidence to reference targets;
Measurement: deriving metrics or signals from execution;
Judging: evaluating execution against criteria or specifications;
Verification: checking explicit constraints or claims;
Task Progress: degree of task advancement;
Execution Quality: quality of the execution process;
Requirement Satisfaction: satisfaction of specified requirements.
}
\\
\hline

\end{tabularx}
\endgroup
\end{table*}

\subsubsection{Agent Execution Assessment}
\label{sec:agent-execution-assessment}

Final outcomes provide only a coarse indication of performance, leaving execution progress, process quality, and requirement satisfaction unclear. Unlike failure diagnosis, which attributes observed failures (see Section~\ref{sec:failure-diagnosis}), execution assessment evaluates recorded behaviors, states, or artifacts against assessment references. Accordingly, we distinguish \textbf{Process Quality Evaluation}, which characterizes progress or quality, from \textbf{Execution Verification}, which checks requirement satisfaction or support for specific claims. Table~\ref{tab:aea-characteristics} summarizes their task-specific characteristics.

\textbf{Process Quality Evaluation.}
Process Quality Evaluation assesses intermediate execution evidence beyond final outcomes. One pattern matches observed states against predefined task references: AgentBoard scores completed human-annotated subgoals, whereas LlamaTouch checks whether task-critical states are reached during mobile-agent execution~\cite{ma2024agentboard,zhang2024llamatouch}. The former captures partial completion, while the latter emphasizes essential states; both depend on how well their references represent legitimate execution paths.

A second pattern applies multiple behavioral or quality criteria. Process Evaluation checks intermediate behaviors against a compliance checklist, while AutoGen-TraceKit derives process and robustness measurements from fine-grained interaction traces~\cite{gritta2026process,mallikarachchi2026autogentracekit}. AgentDiagnose evaluates competencies, whereas Beyond the Final Answer and TRACE assess multiple trajectory-quality dimensions, with TRACE aggregating them through a hierarchical utility formulation~\cite{ou2025agentdiagnose,kim2025beyond,chen2026traceevaluation}. These approaches differ in abstraction and aggregation: behavioral or trace-level measures directly reflect observable evidence, whereas competency or quality dimensions require stronger mappings from logs to abstract criteria. Separate dimensions preserve trade-offs, while aggregation yields a compact overall judgment.

A third pattern decomposes assessment by execution level or decision stage. Agentic CLEAR evaluates records at system, trace, and node levels, whereas Aghzal et al. assess Web-agent planning, execution, and replanning~\cite{yehudai2026agentic,aghzal2026webagentsfail}. Both reveal deficiencies that a single trajectory-level result may obscure. However, they organize assessment by execution-representation level and decision-process stage, respectively.
\textbf{Execution Verification.}
Progress or quality judgments alone do not establish whether specific requirements hold or claims are supported. \textbf{Execution Verification.} therefore checks execution records against explicit or constructed requirements to determine whether those requirements are met or specific claims are supported.

One pattern verifies behavioral or normative requirements. TARV aggregates runtime trust signals from actions, memory transitions, and inter-agent communication into an overall judgment, whereas AudAgent formalizes natural-language privacy policies and checks observed agent data practices against them~\cite{patel2026runtimeverification,zheng2025audagent}. Thus, TARV constructs judgments from execution-derived signals, while AudAgent uses an explicit external policy, preserving a clearer correspondence between observed behavior and the requirement checked.

Another pattern verifies artifacts through provenance-bearing evidence. ProvenanceGuard decomposes answers into factual claims, links them to source-bearing MCP tool outputs, and checks semantic support and source attribution~\cite{alvarez2026provenanceguard}. VET instead binds outputs to agent identity, configuration, and execution provenance to verify their declared execution origin~\cite{grigor2026vet}. Both use provenance, but the former focuses on source-supported factual claims, whereas the latter focuses on identity and execution-provenance authenticity. Such verification establishes specific evidential claims rather than an overall execution-quality judgment.

\begin{tcolorbox}[
    colback=summarybg,
    colframe=black,
    boxrule=0.5pt,
    arc=0pt,
    left=5pt,
    right=5pt,
    top=4pt,
    bottom=4pt
]
\textbf{Summary.}
Agent Execution Assessment extends evaluation beyond final task outcomes through process- and evidence-based judgments of progress, quality, or requirement satisfaction.
\textbf{Process Quality Evaluation} assesses \emph{how well} intermediate execution progresses or performs using task states, behavioral criteria, quality dimensions, or process decompositions, yielding progress scores or quality assessments~\cite{ma2024agentboard,zhang2024llamatouch,ou2025agentdiagnose,chen2026traceevaluation}.
\textbf{Execution Verification} instead determines \emph{whether} recorded behavior or provenance satisfies specified requirements or supports particular claims~\cite{patel2026runtimeverification,zheng2025audagent,alvarez2026provenanceguard,grigor2026vet}.
The two families mainly differ in their \emph{assessment objective}, \emph{reference basis}, and \emph{judgment form}.
The closely related agent trustworthiness assessment instead characterizes security or privacy properties under specified risk conditions (see Section~\ref{sec:agent-trustworthiness-assessment}).
\end{tcolorbox}

\begin{table*}[t]
\centering
\caption{Task-Specific Characteristics of Agent Trustworthiness Assessment Studies}
\label{tab:ata-characteristics}

\begingroup
\normalfont
\fontsize{7.2pt}{8.3pt}\selectfont
\renewcommand{\arraystretch}{1.08}
\setlength{\tabcolsep}{3.2pt}

\begin{tabularx}{0.95\textwidth}{
|>{\centering\arraybackslash}m{0.175\textwidth}
|>{\centering\arraybackslash}m{0.165\textwidth}
|>{\centering\arraybackslash}m{0.15\textwidth}
|>{\centering\arraybackslash}m{0.155\textwidth}
|>{\centering\arraybackslash}X|
}
\hline

\rowcolor{gray!20}
\textbf{Study} &
\textbf{Property} &
\textbf{Structure} &
\textbf{Strategy} &
\textbf{Scenario}
\\
\hline

Ruan et al.~\cite{ruan2024toolemu}
& Security
& Single
& Oracle-Based
& Attack Scenarios
\\
\hline

Zhang et al.~\cite{zhang2025agentsecuritybench}
& Security
& Multi
& Oracle-Based
& Attack Scenarios
\\
\hline

Ying et al.~\cite{ying2025securewebarena}
& Security
& Multi
& Oracle-Based
& Attack Scenarios
\\
\hline

Zhang et al.~\cite{zhang2025mcp}
& Security
& Multi
& Oracle-Based
& Attack Scenarios
\\
\hline

Ersoy et al.~\cite{ersoy2026darkpatterns}
& Integrity
& Single
& Comparative
& Controlled Manipulation
\\
\hline

Atinafu et al.~\cite{atinafu2026rewardhacking}
& Integrity
& Single
& Comparative
& Controlled Manipulation
\\
\hline

Biswas et al.~\cite{biswas2025privacy}
& Privacy
& Single
& Quantification
& Multi-Agent Communication
\\
\hline

El Yagoubi et al.~\cite{yagoubi2026agentleak}
& Privacy
& Single
& Quantification
& Multi-Agent Communication
\\
\hline

Choong et al.~\cite{choong2026capchain}
& Privacy
& Multi
& Policy-Based Control
& Multi-Agent Communication
\\
\hline

\multicolumn{5}{
|>{\raggedright\arraybackslash}p{\dimexpr0.95\textwidth-2\tabcolsep-2\arrayrulewidth\relax}|
}{
Property: trustworthiness property being assessed;
Structure: organization of the assessment process;
Strategy: strategy used to derive the assessment result;
Scenario: evaluation conditions under which the assessment is conducted;
Security: resistance to security threats;
Integrity: robustness to manipulation or evaluation compromise;
Privacy: protection against sensitive-information exposure;
Single: one assessment stage;
Multi: multiple assessment stages;
Oracle-Based: assessment using evaluators or predefined criteria;
Comparative: comparison across controlled conditions;
Quantification: measurement of information exposure;
Policy-Based Control: assessment through policy- or capability-based decisions over sensitive-information propagation;
Attack Scenarios: risk or adversarial conditions;
Controlled Manipulation: deliberately varied manipulation conditions;
Multi-Agent Communication: information exchange among agents.
}
\\
\hline

\end{tabularx}

\endgroup
\end{table*}

\subsubsection{Agent Trustworthiness Assessment}
\label{sec:agent-trustworthiness-assessment}

Agent execution records can also be used to characterize trustworthiness properties that extend beyond the quality or correctness of a single execution. Whereas Agent Execution Assessment evaluates execution progress, quality, or requirement satisfaction, trustworthiness assessment examines how agent systems behave under specified security, manipulation, or privacy conditions. Existing approaches differ mainly in the trustworthiness property being assessed and the conditions under which execution evidence is collected. Accordingly, we organize existing studies into \textbf{Security Risk Assessment} and \textbf{Privacy Leakage Assessment}. Table~\ref{tab:ata-characteristics} summarizes their task-specific characteristics.

\textbf{Security Risk Assessment.}
Security Risk Assessment examines unsafe or integrity-compromising behavior under explicitly constructed threat conditions. One common pattern is to expose agents to predefined attack scenarios and aggregate execution outcomes across the resulting runs. ToolEmu evaluates risky tool use in emulated high-stakes environments, while Agent Security Bench covers attacks across system prompts, user inputs, memory, and tools~\cite{ruan2024toolemu,zhang2025agentsecuritybench}. SecureWebArena and MCP Security Bench instead specialize the evaluation environment to Web-agent workflows and MCP-based interactions, respectively~\cite{ying2025securewebarena,zhang2025mcp}. These approaches share the use of adversarial execution scenarios as the basis for assessment, but differ in threat coverage and execution boundary. Some concentrate on particular tool-use risks, whereas others evaluate multiple attack surfaces or domain-specific interaction structures. Their resulting security profiles therefore depend on which threats and operational contexts are represented in the evaluation setting.

A second pattern uses controlled manipulation to isolate how a particular factor affects agent behavior. Ersoy et al.~\cite{ersoy2026darkpatterns} vary the presence of dark patterns in Web environments and compare their influence on agent actions, whereas Atinafu et al.~\cite{atinafu2026rewardhacking} manipulate evaluator-tampering opportunities, test-information access, and protection settings to examine evaluation integrity. Both construct comparative execution conditions around a specific risk factor, but target different trust boundaries: environmental influence in the former and evaluation-process integrity in the latter. Compared with broader attack-scenario assessment, controlled manipulation narrows the evaluated condition so that behavioral changes can be related more directly to the varied factor.

\textbf{Privacy Leakage Assessment.}
Privacy Leakage Assessment examines sensitive-information exposure and propagation during agent communication, focusing on leakage quantification, channel-level localization, and communication control.

Biswas et al.~\cite{biswas2025privacy} quantify leakage in multi-agent exchanges under probing, membership inference, prompt injection, repeated interaction, and colluding-adversary settings. Their complementary measures enable comparisons across communication and adversarial conditions.

AgentLeak~\cite{yagoubi2026agentleak} distinguishes final outputs from internal inter-agent channels and compares exposure across communication paths. This channel-level analysis captures sensitive-information propagation through internal messages that output-only inspection may miss.

CapChain~\cite{choong2026capchain} combines privacy analysis with communication control by checking messages against attribute--purpose constraints at the state-transition layer, redacting fields that violate these constraints, and recording decisions with provenance. Its results therefore characterize both communication behavior and privacy-control decisions; reported redaction activations should not be interpreted directly as the amount of privacy leakage prevented.

\begin{tcolorbox}[
    colback=summarybg,
    colframe=black,
    boxrule=0.5pt,
    arc=0pt,
    left=5pt,
    right=5pt,
    top=4pt,
    bottom=4pt
]
\textbf{Summary.}
Agent Trustworthiness Assessment extends agent log mining by using execution evidence to characterize security or privacy properties under specified risk conditions.
\textbf{Security Risk Assessment} examines unsafe or integrity-compromising behavior through attack scenarios or controlled manipulations, with assessments differing in threat coverage and execution conditions~\cite{ruan2024toolemu,zhang2025agentsecuritybench,ying2025securewebarena,ersoy2026darkpatterns}.
\textbf{Privacy Leakage Assessment} instead analyzes sensitive-information exposure and propagation across agent communication by measuring leakage, localizing affected channels, or evaluating communication-control decisions~\cite{biswas2025privacy,yagoubi2026agentleak,choong2026capchain}.
The two families differ mainly in their \emph{trustworthiness property}, \emph{evaluation condition}, and \emph{evidence focus}, with assessments remaining conditional on the evaluated threat model, communication setting, and assessment criteria.
The closely related agent execution assessment instead focuses on process quality or requirement satisfaction (see Section~\ref{sec:agent-execution-assessment}).
\end{tcolorbox}

\subsection{Discussion}
\label{sec:discussion}

\textbf{Finding 8: Analytical Reduction.}
\textbf{\emph{Long-trajectory mining increasingly relies on reducing execution evidence or candidate spaces, but premature reduction may discard information required by the target analysis.}}
Existing studies reduce analytical complexity in different ways. In Execution Anomaly Detection, long trajectories are reorganized or selectively filtered before judgment~\cite{kale2025reliable,arike2026does}, whereas Failure Diagnosis narrows competing hypotheses, execution scopes, or failure types before attribution~\cite{zhu2026raffles,sun2026scope,qiao2026verifymas}. These strategies make long-trajectory analysis more tractable, but the reduction step also determines what evidence remains available afterward. If a decisive trajectory segment or the actual failure candidate is removed too early, subsequent analysis cannot recover it. Future work should therefore reduce irrelevant evidence and candidate spaces while preserving information necessary for the intended analytical objective.

\textbf{Finding 9: Analytical Unit.}
\textbf{\emph{Entity-level units such as executions, agents, or steps are insufficient for some problems that emerge from interactions, dependencies, or communication processes.}}
Existing mining approaches often assign results to individual execution entities, but several studies show that the phenomenon being analyzed may arise from relations among them. MAStitch combines local agent analysis with global aggregation to expose cross-agent anomalies~\cite{vaknin2026mastitch}; Who\&When separates responsible-agent and decisive-step attribution, while StepFinder models temporal and cross-agent dependencies among execution steps~\cite{zhang2025agentcauses,zhu2026stepfinder}. Similarly, AgentLeak shows that privacy exposure may occur through internal communication channels even when it is absent from final outputs~\cite{yagoubi2026agentleak}. Thus, the unit where a problem emerges does not necessarily coincide with the entity receiving the final analytical label. Relational units such as interactions, dependency paths, handoffs, and communication channels deserve further attention.

\textbf{Finding 10: Analytical Depth.}
\textbf{\emph{The analytical depth required by agent log mining depends on the analytical objective, and finer-grained analysis is not inherently preferable.}}
Execution Anomaly Detection may only require a reliable abnormality judgment, although some approaches further localize irregular behavior to agents or tool actions~\cite{vaknin2026mastitch,mou2026toolsafe}. Failure Diagnosis generally requires deeper analysis, ranging from responsibility localization to explanation, remediation support, and intervention-based examination~\cite{zhang2025agentcauses,mulian2026agentfixer,ma2026dover}. Agent Execution Assessment similarly ranges from progress measurement to criterion-level or multi-level quality judgments~\cite{ma2024agentboard,chen2026traceevaluation,yehudai2026agentic}. These outputs serve different purposes and should not be treated as interchangeable refinements of the same result. Future work should clarify what level of detection, localization, explanation, or validation is sufficient for each analytical objective.

\textbf{Finding 11: Judgment Rule.}
\textbf{\emph{Agent-log-based judgments are inherently dependent on the reference used to interpret execution evidence, and no single judgment rule is appropriate for all analytical properties.}}
Execution assessment illustrates this dependence clearly: progress can be judged against subgoals or essential states, while process quality may rely on behavioral criteria, competency dimensions, or higher-level rubrics~\cite{ma2024agentboard,zhang2024llamatouch,ou2025agentdiagnose,chen2026traceevaluation}. Verification approaches instead use policies, source-support relations, or provenance claims~\cite{zheng2025audagent,alvarez2026provenanceguard,grigor2026vet}, while trustworthiness assessments depend on specified threat or communication conditions~\cite{ruan2024toolemu,biswas2025privacy}. Consequently, the same execution may satisfy one set of assessment criteria while violating another. Assessment along broader dimensions generally requires more semantic interpretation of concrete execution evidence. Future work should make the applicability and coverage of such judgment rules explicit rather than treating analytical scores as context-independent properties.

\textbf{Finding 12: Failure Evolution.}
\textbf{\emph{Failure localization identifies where responsibility may lie, but existing diagnosis still provides limited explanation of how an originating error evolves into the final observed failure.}}
Who\&When already shows that the responsible agent and decisive error step may differ~\cite{zhang2025agentcauses}, while ERRORPROBE and StepFinder introduce backward or dependency-aware analysis to connect observed symptoms with earlier execution evidence~\cite{li2026self,zhu2026stepfinder}. These approaches improve localization, but an early reasoning or decision error may subsequently be transformed through plans, inter-agent messages, tool actions, and execution states before the final failure appears. Downstream behavior may also propagate, amplify, or partially compensate for the original problem. Thus, failure formation is not necessarily a point-to-point relation between one erroneous step and one final symptom. Future diagnosis should further explain how errors propagate and transform throughout execution rather than stopping at isolated root-cause localization.

\textbf{Finding 13: Interventional Analysis.}
\textbf{\emph{Interventional analysis remains relatively limited even when the analytical objective requires determining whether changing a suspected execution factor affects the final outcome.}}
Most attribution approaches infer responsibility from observed trajectories through hypothesis verification, learned scoring, cross-execution comparison, or historical evidence~\cite{qiao2026verifymas,geng2026failurelocalization,ge2026spectrum}. Such evidence can identify plausible failure sources, but association alone does not show whether modifying the candidate would alter the outcome. DoVer directly intervenes on suspected messages, instructions, or plans and re-executes subsequent workflow stages~\cite{ma2026dover}, while controlled replay similarly allows localized modifications to be examined with preceding execution evidence preserved~\cite{luan2026repair}. Controlled manipulation also appears in trustworthiness assessment~\cite{ersoy2026darkpatterns,atinafu2026rewardhacking}. These approaches provide stronger evidence about outcome sensitivity, but require executions or environments that can be modified and replayed. Their applicability may therefore be limited by irreversible actions, external dependencies, or execution cost, motivating further study of when interventional analysis is feasible and necessary.

\section{Open-source Toolkits and Datasets}
\label{sec:Open-source-Toolkits-and-Datasets}

A growing number of open-source tools and public datasets are available for agent log analysis. We include publicly reusable tools that directly capture or process agent-generated logs, traces, or trajectories and support at least one stage of the analysis pipeline, excluding closed-source commercial services such as LangSmith~\cite{langsmith_commercial} and Arize AX~\cite{arize_ax_commercial}. Table~\ref{tab:agent_log_resources} summarizes these resources, which are discussed according to their roles in Agent Logging, Agent Log Structuring, and Agent Log Mining, followed by representative public datasets.

\begin{table*}[p]
\centering

\rotatebox{90}{%
\begin{minipage}{0.95\textheight}
\centering

\caption{A List of Open-source Tools and Datasets for Agent Log Analysis}
\label{tab:agent_log_resources}

\normalfont
\fontsize{7.2pt}{8.3pt}\selectfont
\renewcommand{\arraystretch}{1.08}
\setlength{\tabcolsep}{3.2pt}

\begin{tabularx}{\linewidth}{
@{}|
>{\centering\arraybackslash}m{0.2cm}|
>{\centering\arraybackslash}m{2.6cm}|
>{\centering\arraybackslash}m{0.65cm}|
>{\centering\arraybackslash}m{1.05cm}|
>{\arraybackslash}X|
>{\centering\arraybackslash}m{0.65cm}
@{}|
}

\hline
\textbf{} &
\textbf{Resource} &
\textbf{Year} &
\textbf{Agent Setting} &
\multicolumn{1}{c|}{\textbf{Description}} &
\textbf{Link}
\\
\hline


\multirow{20}{*}{\rotatebox[origin=c]{270}{\textbf{Tools}}}

& MLflow
& 2024
& Both
& An observability platform for tracing, evaluating, and monitoring LLM and agent applications.
& \cite{mlflow_tool} \\

& OpenInference
& 2023
& Both
& An OpenTelemetry framework for standardized AI tracing.
& \cite{openinference_tool} \\

& OpenLLMetry
& 2023
& Both
& An OpenTelemetry-based observability toolkit for tracing and monitoring LLM and agent applications.
& \cite{openllmetry_tool} \\

& OpenLIT
& 2024
& Both
& An observability platform for collecting and analyzing traces, metrics, and logs from AI applications.
& \cite{openlit_tool} \\

& AgentOps
& 2024
& Both
& An observability tool for tracing, monitoring, and replaying agent executions.
& \cite{agentops_tool} \\

& Arize Phoenix
& 2023
& Both
& An observability tool for tracing, evaluating, and troubleshooting agent applications.
& \cite{phoenixTracingDocs} \\

& Langfuse
& 2023
& Both
& An observability tool for tracing, inspecting, and evaluating LLM and agent executions.
& \cite{langfuseObservabilityDocs} \\

& Opik
& 2024
& Both
& An observability tool for tracing, evaluating, and monitoring agent executions.
& \cite{opik_tool} \\

& TruLens
& 2023
& Both
& An evaluation tool for tracing and assessing agent execution processes.
& \cite{trulens_tool} \\

& AgentSight
& 2025
& Both
& A system-level observability tool for agent execution correlation.
& \cite{agentsight_tool} \\

& ReDel
& 2024
& Multi
& An event-driven logging and replay tool for recursive multi-agent executions.
& \cite{redel_tool} \\

& AgentsView
& 2026
& Single
& A local-first analysis tool for importing, searching, and inspecting coding-agent session logs.
& \cite{agentsview_tool} \\

& VCC
& 2026
& Both
& A trace compilation tool for transforming raw agent conversation logs into structured views.
& \cite{vcc_tool} \\

& Graph of Trace
& 2026
& Single
& A trace structuring tool for organizing agent execution events into verifiable dependency graphs.
& \cite{graphoftrace_tool} \\

& AGDebugger
& 2025
& Multi
& An interactive debugging tool for inspecting, editing, and replaying multi-agent execution histories.
& \cite{agdebugger_tool} \\

& AgentRx
& 2026
& Both
& A trace diagnosis toolkit for normalizing agent traces, localizing failure steps, and classifying root causes.
& \cite{agentrx_tool} \\

& AgentEvals
& 2025
& Single
& A trajectory evaluation toolkit for assessing agent trajectories, intermediate steps, and tool calls.
& \cite{agentevals_tool} \\

& Hodoscope
& 2026
& Single
& An interactive trajectory analysis toolkit for visualizing and exploring behavioral patterns.
& \cite{hodoscope_tool} \\

& CodeTracer
& 2026
& Single
& A trajectory analysis tool for structuring agent runs and diagnosing step-level execution failures.
& \cite{codetracer_tool} \\

& AgentDebug
& 2025
& Single
& A trajectory diagnosis tool for detecting, localizing, and recovering from agent execution failures.
& \cite{agentdebug_tool} \\

\hline


\multirow{18}{*}{\rotatebox[origin=c]{270}{\textbf{Datasets}}}

& SWE-chat
& 2026
& Single
& A real-world coding-agent session dataset for interaction analysis.
& \cite{swechat_dataset} \\

& TRAIL
& 2025
& Mixed
& An error-annotated agent execution-trace dataset for reasoning analysis and failure localization.
& \cite{trail_dataset} \\

& TraceSafe-Bench
& 2026
& Single
& A tool-calling trajectory dataset for analyzing functional errors and security-risk behaviors.
& \cite{tracesafe_dataset} \\

& AgentRx
& 2026
& Mixed
& An agent failure-log dataset for failure localization and root-cause analysis across agent systems.
& \cite{agentrx_dataset} \\

& TraceElephant
& 2026
& Mixed
& An agent execution-log dataset for failure diagnosis.
& \cite{traceelephant_dataset} \\

& Who \& When
& 2025
& Multi
& A multi-agent failure-trace dataset for responsible-agent attribution and error-step localization.
& \cite{whoandwhen_dataset} \\

& CoderForge-Preview
& 2026
& Single
& A long-horizon coding-agent trajectory dataset for repository-level software-engineering tasks.
& \cite{coderforge_dataset} \\

& MAST-Data (MAD)
& 2025
& Multi
& A multi-agent execution dataset for failure-mode analysis.
& \cite{mast_dataset} \\

& AEGIS
& 2025
& Multi
& A controlled-error multi-agent trajectory dataset for faulty-agent and error-mode attribution.
& \cite{aegis_dataset} \\

& AgentProcessBench
& 2026
& Single
& A tool-using agent trajectory dataset for step-wise process quality evaluation.
& \cite{agentprocessbench_dataset} \\

& CodeTraceBench
& 2026
& Single
& A coding-agent trajectory dataset for step-level execution error diagnosis.
& \cite{codetracebench_dataset} \\

& TELBench
& 2026
& Single
& A deep-research agent trajectory dataset for span-level execution error localization.
& \cite{telbench_dataset} \\

& AgentErrorBench
& 2025
& Single
& An annotated agent failure dataset for root-cause diagnosis across diverse agent environments.
& \cite{agenterrorbench_dataset} \\

& Exgentic
& 2026
& Single
& An OpenTelemetry-shaped chat-trace dataset for replay and behavioral analysis of agent executions.
& \cite{exgentic_traces_dataset} \\

& SWE-agent-trajectories
& 2024
& Single
& A large-scale coding-agent trajectory dataset for analyzing reasoning, actions, observations, and patches.
& \cite{sweagent_trajectories_dataset} \\

& SWE-rebench
& 2025
& Single
& An OpenHands trajectory dataset for analyzing multi-turn software-engineering agent executions.
& \cite{swerebench_openhands_dataset} \\

& Terminal-Bench
& 2026
& Single
& A terminal-agent trajectory dataset for analyzing messages, tool calls, and observations.
& \cite{terminalbench_trajectories_dataset} \\

& OSWorld-Verified
& 2025
& Single
& A computer-use agent trajectory dataset for execution analysis.
& \cite{osworld_trajectories_dataset} \\

\hline
\end{tabularx}

\vspace{2pt}
\parbox{\linewidth}{\scriptsize
\textit{Note:} Year denotes the first public release/publication year. Agent Setting denotes supported agent types for tools and log-source types for datasets; Both applies to tools, while Mixed denotes datasets containing both single- and multi-agent logs.
}

\end{minipage}%
}

\end{table*}

\subsection{Open-source Tools}
\label{sec:opensource_tools}

A large share of existing open-source tools primarily supports the Agent Logging stage, with emphasis on runtime tracing, telemetry collection, and observability for LLM and agent applications (e.g., MLflow~\cite{mlflow_tool}, AgentOps~\cite{dong2024agentops}, Arize Phoenix~\cite{phoenixTracingDocs}, and Langfuse~\cite{langfuseObservabilityDocs}).
Other tools further support standardized tracing, framework-integrated instrumentation, or observation of system-level and multi-agent executions (e.g., OpenInference~\cite{openinference_tool}, OpenLLMetry~\cite{openllmetry_tool}, AgentSight~\cite{zheng2025agentsight}, and ReDel~\cite{zhu2024redel}).
Overall, open-source support is most mature at the logging stage, reflecting the practical demand for making agent executions observable before downstream analysis.

Fewer tools primarily target Agent Log Structuring.
These tools organize recorded executions into searchable, structured, or graph-based representations to facilitate retrieval, inspection, and navigation (e.g., AgentsView~\cite{agentsview_tool}, VCC~\cite{zhang2026vcc}, Graph of Trace~\cite{gao2026graphtrace}, and AGDebugger~\cite{epperson2025interactive}).
Compared with logging-oriented tooling, structuring approaches remain more heterogeneous and have not yet converged on a common intermediate representation.

At the Agent Log Mining stage, open-source tools mainly support execution evaluation, behavioral analysis, failure localization, and diagnosis (e.g., AgentEvals~\cite{agentevals_tool}, Hodoscope~\cite{hodoscope_tool}, AgentRx~\cite{agentrx_tool}, and AgentDebug~\cite{agentdebug_tool}).
These tools are generally more task-specific than logging-oriented platforms, with their designs centered on particular analytical objectives such as trajectory assessment or failure diagnosis.

\subsection{Public Datasets}
\label{sec:public_datasets}

Besides open-source tools, public datasets provide execution records from diverse agent systems and task environments for agent log analysis.
As summarized in Table~\ref{tab:agent_log_resources}, these datasets vary in their collection settings, ranging from traces obtained from actual agent executions to trajectories generated or modified under controlled conditions.
A substantial group of datasets focuses on execution failures, supporting tasks such as failure localization, responsible-agent attribution, and error diagnosis across single- and multi-agent systems (e.g., TRAIL~\cite{trail_dataset}, Who \& When~\cite{whoandwhen_dataset}, MAST-Data~\cite{cemri2025do}, and AEGIS~\cite{kong2026aegis}).
Some datasets further target functional or security-related failures in tool-using agents (e.g., TraceSafe-Bench~\cite{tracesafe_dataset}). 

Other datasets capture more general agent behaviors and long-horizon executions, particularly in software-engineering environments (e.g., SWE-chat~\cite{swechat_dataset} and CoderForge-Preview~\cite{coderforge_dataset}).
Overall, existing datasets cover increasingly diverse agent settings and analytical objectives, but remain heterogeneous in execution format, annotation granularity, and data-collection methodology.

\section{Challenges and Opportunities}
\label{sec:challenges-and-opportunities}

\subsection{Existing Challenges}
\label{sec:existing-challenges}

\subsubsection{Task-Aligned Evidence Sufficiency Problem}
\label{subsec:challenge_evidence_sufficiency}

Reliable agent log analysis depends on whether the recorded evidence is sufficient for the intended analytical objective. Existing logging approaches may provide detailed local execution records while still missing interactions and dependencies across agents or system components (see \textbf{Finding 1}). They may also preserve rich operational information without equally complete reasoning- or permission-related semantics (see \textbf{Finding 2}). Consequently, richer traces do not necessarily provide the evidence required by every downstream task, and evidence omitted during execution cannot be faithfully recovered by later structuring or mining. Evidence sufficiency is therefore task-dependent rather than a function of trace size or field count. The central challenge is to determine the minimal yet sufficient execution evidence required for different analytical objectives.

\subsubsection{Cross-Framework Semantic Portability Problem}
\label{subsec:challenge_semantic_portability}

Cross-framework reuse requires not only compatible trace formats but also comparable execution semantics. Runtime-integrated instrumentation preserves agent-specific context but remains coupled to framework-specific execution mechanisms, whereas external interception reduces such coupling at the cost of weaker internal context (see \textbf{Finding 3}). Structuring can further normalize heterogeneous records into common forms, but structurally similar elements may still represent different execution semantics across sources (see \textbf{Finding 5}). Heterogeneity is therefore often shifted from trace format to adapters, mappings, and semantic conventions rather than eliminated. The central challenge is to preserve consistent execution meaning across frameworks, capture mechanisms, and evolving source-specific adaptations.

\subsubsection{Structured Evidence Validity Problem}
\label{subsec:challenge_structured_evidence_validity}

Agent log structuring increasingly produces analytical objects that are constructed rather than directly observed. Schema normalization mainly reorganizes recorded information, whereas semantic or relation construction may introduce boundaries, labels, or inferred content that were not explicitly emitted during execution (see \textbf{Finding 6}). A structured artifact may therefore be useful for downstream analysis without being equivalent to a historical account of the original execution. Source linkage can establish where an artifact originates, but does not by itself establish that its transformation is faithful, complete, or stable. The central challenge is to make the source support and transformation validity of structured evidence explicit before it is used by downstream analysis.

\subsubsection{Multi-Level Execution Coherence Problem}
\label{subsec:challenge_execution_coherence}

Agent execution can be represented at multiple levels, from local events and invocations to semantic artifacts and global execution relations, yet correspondence across these levels remains limited (see \textbf{Finding 4}). Relation-oriented representations make dependencies explicit, but local messages, arguments, observations, and states are often retained at a different granularity from the global structure (see \textbf{Finding 7}). As a result, a high-level analytical conclusion may be difficult to trace back to the local evidence that supports it, while conclusions produced at different granularities may not refer to the same execution phenomenon. The central challenge is to maintain explicit structural and semantic correspondence across local evidence, semantic abstractions, and global execution relations.

\subsubsection{Adaptive Analytical Formulation Problem}
\label{subsec:challenge_analytical_formulation}

Agent log mining must determine what evidence to retain, which analytical unit to model, and how broadly to define the analysis scope. Evidence reduction and candidate restriction improve tractability, but premature reduction may discard information that later becomes decisive (see \textbf{Finding 8}). Analytical units also cannot always be fixed to executions, agents, or steps because some phenomena emerge from interactions and dependency paths (see \textbf{Finding 9}). Failure evolution further changes which part of an execution is analytically relevant, since an originating error may propagate before appearing as a later symptom (see \textbf{Finding 12}). The central challenge is therefore to adapt evidence scope, analytical units, and candidate boundaries jointly to the analytical objective and the evolving execution condition.

\subsubsection{Analytical Claim Validity Problem}
\label{subsec:challenge_claim_validity}

Agent log mining produces outputs with different analytical depths, but finer-grained results do not automatically support stronger claims (see \textbf{Finding 10}). Their interpretation also depends on the reference basis used for judgment, such as behavioral criteria, policies, provenance claims, or threat conditions (see \textbf{Finding 11}). In failure diagnosis, responsibility may remain ambiguous because an originating error and its downstream manifestations can both be associated with the final failure (see \textbf{Finding 12}). More fundamentally, observational attribution can identify plausible failure sources, but does not establish that modifying the suspected factor would change the outcome; stronger causal claims require interventional evidence when such intervention is feasible. The central challenge is therefore to align the semantics and strength of analytical claims with the evidence, reference basis, and validation regime that actually support them.
\subsection{Future Opportunities}
\label{sec:future-opportunities}

\subsubsection{Task-Aware Evidence Specification}
\label{sec:opportunity-evidence-specification}

Addressing the Task-Aligned Evidence Sufficiency Problem requires moving from framework-defined logging schemas toward task-aware evidence specification. Future work should explicitly characterize the observation scope, semantic fields, and granularity required by different analytical objectives. Evidence sensitivity can further be studied through controlled removal or withholding of record types to distinguish essential, conditional, and redundant evidence. Based on such analyses, researchers can seek minimal sufficient evidence sets rather than assuming that richer telemetry is always preferable. This would provide a principled basis for balancing analytical sufficiency against unnecessary capture, storage, and exposure.

\subsubsection{Adaptive Multi-Boundary Evidence Capture}
\label{sec:opportunity-multiboundary-capture}

Evidence requirements must also be translated into effective capture across heterogeneous runtime boundaries. Existing work shows that runtime-integrated instrumentation and external interception provide complementary visibility into agent execution (e.g., AgentTrace, AgentSight, and OpenTelemetry-based approaches~\cite{alsayyad2026agenttrace,zheng2025agentsight,seyedghorban2026observability}). Future logging systems could therefore select observation boundaries according to task-specific evidence requirements rather than relying on one fixed mechanism. Evidence from different boundaries should be correlated through stable execution identities, while capture granularity could increase dynamically when suspicious behavior is detected. Such adaptive multi-boundary capture could improve evidence coverage while reducing dependence on any single framework or runtime layer.

\subsubsection{Cross-Framework Semantic Contracts}
\label{sec:opportunity-semantic-contracts}

The Cross-Framework Semantic Portability Problem calls for interoperability beyond common data formats. Existing intermediate representations demonstrate that heterogeneous trajectories can be mapped into shared structures (e.g., ADP~\cite{song2026adp} and VCC~\cite{zhang2026vcc}), but future work should make their execution semantics explicit through cross-framework semantic contracts. Such contracts could define invariant meanings for core concepts such as invocation, message, delegation, state transition, identity, ordering, and parent--child relations, while preserving framework-specific extensions. Versioning and compatibility rules are also needed as frameworks and telemetry interfaces evolve. This would make cross-framework analysis depend on explicit semantic guarantees rather than hidden assumptions in adapters and converters.

\subsubsection{Verifiable and Provenance-Aware Structuring}
\label{sec:opportunity-verifiable-structuring}

Addressing the Structured Evidence Validity Problem requires structured artifacts to expose how they were derived from execution records. Existing approaches already provide useful foundations through source-preserving mappings, semantic reconstruction, and evidence-oriented representations (e.g., VCC~\cite{zhang2026vcc}, Watson~\cite{rombaut2025watson}, and related provenance-oriented methods~\cite{zhuang2026trajectories}). Future representations should distinguish directly observed, normalized, summarized, annotated, and inferred elements and retain the source records and transformations supporting each artifact. Model- or human-assisted construction should additionally preserve uncertainty or alternative interpretations when appropriate. Verification criteria should then be matched to artifact type, such as preservation for normalization, source support for semantic artifacts, and provenance for inferred relations. This would allow downstream analysis to reason explicitly about evidence with different epistemic strengths.

\subsubsection{Hierarchical Multi-Level Execution Representations}
\label{sec:opportunity-multilevel-representations}

The Multi-Level Execution Coherence Problem suggests a need for representations that connect local execution evidence with higher-level semantic and relational structures. Existing systems already construct provenance, workflow, and behavioral abstractions from agent traces (e.g., Graph of Trace~\cite{gao2026graphtrace}, AgentLens~\cite{lu2025agentlens}, and AgentGraph~\cite{wu2026agentgraph}). Future work should organize these views hierarchically, with explicit mappings from higher-level units and relations back to the lower-level events that support them. Bidirectional mappings are particularly important for tracing global patterns to decisive source evidence and for understanding how individual events contribute to multiple higher-level structures. Such representations would allow different analytical tasks to operate at appropriate granularities without losing correspondence with the same underlying execution.

\subsubsection{Adaptive and Evolution-Aware Analytical Formulation}
\label{sec:opportunity-adaptive-formulation}

Addressing the Adaptive Analytical Formulation Problem requires evidence scope, analytical units, and candidate boundaries to evolve during analysis rather than remain fixed. Existing approaches already demonstrate the value of candidate restriction and suspiciousness ranking (e.g., graph-based diagnosis and FAMAS~\cite{wang2026agenttracerca,ge2026spectrum}). Future methods could begin with compact evidence regions and restore excluded records when later observations invalidate the current hypothesis. Analytical units should also shift between entities and relations when failures arise from coordination, information flow, or dependency paths rather than isolated agents or steps. Candidate spaces could similarly expand or contract as originating errors, propagation effects, and compensating actions become distinguishable. Joint adaptation of these analytical choices would make long-horizon analysis more tractable without permanently discarding potentially decisive evidence.

\subsubsection{Claim-Aligned Evaluation Protocols}
\label{sec:opportunity-claim-evaluation}

The Analytical Claim Validity Problem requires evaluation protocols to reflect what an analysis actually claims. Existing assessment methods already employ different reference bases and analytical units (e.g.,  LlamaTouch~\cite{zhang2024llamatouch}, Process Evaluation~\cite{gritta2026process}, and AudAgent~\cite{zheng2025audagent}), highlighting that their outputs are not directly interchangeable. Future studies should therefore specify the semantic type, analytical unit, and depth of each target claim and construct ground truth at the same level. Evaluation should also permit ambiguous or multiple responsible units when the execution does not support a unique label. Reference quality should be assessed separately from the judging mechanism, and metrics should match the intended claim, distinguishing detection, localization, explanation, reference agreement, profile coverage, and causal validation. Such protocols would make heterogeneous agent-log mining results more interpretable and comparable.

\subsubsection{Intervention-Supported Causal Analysis}
\label{sec:opportunity-causal-analysis}

For causal diagnosis, observational attribution should be treated as a hypothesis rather than the endpoint of analysis. Existing intervention-based methods show that suspected failure sources can be tested through targeted modification and re-execution (e.g., DoVer~\cite{ma2026dover}). Future work should develop controlled replay and state-restoration mechanisms that support interventions on specific messages, plans, tool inputs, states, or agent decisions. Because agent behavior is non-deterministic, causal validation should rely on repeated interventions or alternative execution branches rather than a single successful replay. Evaluation should consequently report outcome change, reproducibility, intervention success, and additional execution cost alongside attribution accuracy. This would provide stronger evidence for distinguishing factors merely associated with failure from those with genuine counterfactual relevance.

\section{Conclusion}
\label{sec:conclusion}

This survey organizes recent research on automated agent log analysis into three stages: Agent Logging, Agent Log Structuring, and Agent Log Mining. It provides a lifecycle-oriented view of how agent-generated execution records are captured, structured, and analyzed. We compare existing approaches using stage-specific criteria and taxonomies, synthesize cross-study findings on their capabilities and limitations, and summarize representative open-source tools and public datasets. Based on these findings, we identify challenges and future research opportunities concerning evidence sufficiency, semantic portability, structured-evidence validity, execution coherence, adaptive analysis, and analytical claim validity.
\bibliographystyle{ACM-Reference-Format}
\bibliography{ref}

\appendix
\end{document}